\documentclass[aps,pra,superscriptaddress,showpacs,nofootinbib,floatfix,twocolumn]{revtex4}

\usepackage{graphicx}	

\newcommand{\beq}{\begin{equation}}
\newcommand{\eeq}{\end{equation}}
\newcommand{\beqa}{\begin{eqnarray}}
\newcommand{\eeqa}{\end{eqnarray}}
\newcommand{\bpr}{\begin{problem}}
\newcommand{\epr}{\end{problem}}
\newcommand{\bcent}{\begin{center}}
\newcommand{\ecent}{\end{center}}
\newcommand{\bfig}{\begin{figure}}
\newcommand{\efig}{\end{figure}}
\newcommand{\bpc}{\begin{picture}}
\newcommand{\epc}{\end{picture}}
\newcommand{\barr}{\begin{array}}
\newcommand{\earr}{\end{array}}
\newcommand{\bitm}{\begin{itemize}}
\newcommand{\eitm}{\end{itemize}}
\newcommand{\bright}{\begin{flushright}}
\newcommand{\eright}{\end{flushright}}
\newcommand{\bminip}{\begin{minipage}}
\newcommand{\eminip}{\end{minipage}}
\newcommand{\btab}{\begin{tabular}}
\newcommand{\etab}{\end{tabular}}

\newcommand{\nnb}{\nonumber}
\newcommand{\reflef}{(\ref}

\newcommand{\hiroshima}{Graduate School of Science, Hiroshima University, Kagamiyama, Higashi-Hiroshima 739-8526, Japan}
\newcommand{\lmu}{Ludwig-Maximilians-Universit$\ddot{a}$t M$\ddot{u}$nchen, Fakult$\ddot{a}$t f. Physik, Am Coulombwall 1, D-85748 Garching, Germany}

\begin{document}

\title{Probing semi-macroscopic vacua by high fields of lasers}

\author{K. Homma} \affiliation{\hiroshima} \affiliation{\lmu}
\author{D. Habs} \affiliation{\lmu}
\author{T. Tajima} \affiliation{\lmu}

\date{\today}

\begin{abstract}
The invention of laser immediately enabled us to detect nonlinearities
of photon interaction in matter, as manifested for example by Franken et al.'s
detection of second harmonic generation and the excitation of the Brillouin
forward scattering process. 
With the recent advancement in high power high energy laser 
and the examples of the nonlinearity study of laser-matter interaction
by virtue of properly arranging laser and detectors,
we envision the possibility of probing nonlinearities of photon
interaction in vacuum over substantial spacetime scales compared with the
microscopic scale provided by high energy accelerators.
The hithertofore never detected Euler-Heisenberg nonlinearities 
in quantum electrodynamics (QED) in vacuum should come within our reach 
of detection using intense laser fields.
Also our method should put us in a position with a far greater sensitivity
of probing possible light-mass fields that have been postulated.
With the availability of a large number of coherent photons our suggested
measurement methods include the phase sensitive (contrast) imaging that
avoids the pedestal noise and the scheme of second harmonic detection of
photon nonlinearities in vacuum over a long co-propagating distance
incurring resonance excitation. 
These methods carve out a substantial swath of new experimental parameter 
regimes of the exploration of photon nonlinearities in vacuum covering 
the force range from the electron mass scale to below neV.
\end{abstract}

\maketitle
\section{Introduction}\label{Intro}
\begin{figure}
\includegraphics[width=1.0\linewidth]{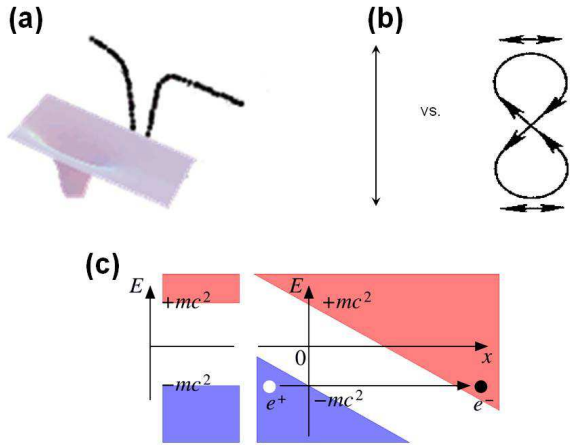}
\caption{
Photon nonlinearities in media.
(a) The nonlinearity arising from atomic anharmonic fields in high intensities;
(b) The relativistic nonlinearity of plasma electrons in relativistically
intense laser makes the harmonic oscillations of electrons~(left) turn into
anharmonic motions~(right);
(c) Even in vacuum photon nonlinearities occur when the field intensity
approaches the Schwinger field, at which the intense field can polarize
the vacuum anharmonically and eventually help virtual electrons and positrons
turn into real particles, a clear manifestation of the breakdown of harmonic
photon fields in the low intensity ordinary conditions.
}
\label{Fig0}
\end{figure}

The invention of laser in 1960 constituted an introduction of the possibility
of coherent intense photon fields at the optical wavelengths.
In fact immediately following it Franken et al.\cite{Franken} observed the
nonlinearity of the quartz crystal generated second harmonics of the laser at
the 'high field' of $10^5$~V/cm. Many other nonlinearities of the media due to
the high intensity of the laser have been discovered by subsequent years,
as compiled in the late 1960's \cite{Bloembergen}. Above the field of
manifestation of the nonlinearity of the second harmonic generation
that Franken et al. encountered, neutral
atoms may be directly ionized at the field called the Keldysh field
\cite{Keldysh}, which is on the order of $10^8$~V/cm ({\it i.e.}
an atomic scale of an eV potential over an \AA~length),
depending on the material. The material's nonlinearities arise from its
polarization under a large enough field of laser beyond the linear atomic
field strength. See Fig.\ref{Fig0}(a).
Typically the restoring force of the electron to the rest of
the atom saturates and can no longer match so much stronger applied field.
Relativistically strong lasers ({\it i.e.} the laser field is so strong
to bring electrons to relativistic energies in an optical cycle) can
induce the relativistic nonlinearity in electron dynamics of plasma,
which is instrumental in producing intense wakefields~\cite{Tajima}.
This is shown in Fig.\ref{Fig0}(b).
Spurred by the promise and requirement of intense fields that can be sustained
in a plasma wakefield excitation,
since the invention of the CPA (chirped pulse amplification)
\cite{CPA} the achievable laser intensity has been exponentially multiplying
\cite{MourouRMP}. Along with this advance of most intense laser
development comes a technique to further the available laser field such as
in \cite{Bulanov2003}, which in principle sees the possibility of
reaching even the Schwinger field characterized by a quantum electrodynamic
(QED) scale of an MeV (an electron-positron pair creation energy)
over the Compton length, {\it i.e.}~$10^{16}$~V/cm.
Just as the material's ionization happens at the Keldysh
field, the vacuum breakdown happens at the Schwinger field \cite{Schwinger}.
This is depicted in Fig.\ref{Fig0}(c).
Even before reaching this Schwinger field a host of nonlinear behaviors under
intense fields are expected \cite{MarklundRMP}.

The study of matter and vacuum behaviors under intense fields these days is
called high field science and reviews may be found in
\cite{MourouRMP, MarklundRMP, TajimaHighFieldScience, TajimaEPJD}.
In the following we suggest a class of possible investigations of vacuum
in high field science, where we take advantage of the large amplitude of
the (laser) electromagnetic fields and the relatively macroscopic scale of
the field spatial scale (typically micron) compared with the microscopic
collider experiments (typically fm or less).
The most successful tool for exploring the fundamental
nature in smallest structures of matter has been that of collider, in which
high energy (and thus high momentum) charged particles are produced
and collide each other to probe the smallest scale with highest energies
($\delta x \sim \hbar/p$). The present approach we suggest here
may be contrasted to this high momentum approach,
in that it explores semi-macro spatial scales and
relatively lower energy processes with high amplitude (or fields) by
{\it 'exciting the constituent matter (or vacuum) and its structure}'.

We may wish to once again borrow the parallelism of the two alternative
investigation methods of atomic physics by particle collisions and by laser
nonlinear optics. The former explores the atom via Rutherford's method of
the high energy beam streaming through and thereby scattered by the constituent
matter deep in its core. By this Rutherford discovered that the ordinary matter
is composed of electrons that pervade the unit of the constituent matter, atom,
while there exists a tiny core, the nucleus, that sharply scatters the incoming
beam. The analysis of the scattering of the injected beam reveals the inner
core property. Ever since this epoch making experiment, this approach paved
the most successful way of the modern particle physics experimental method,
including the collider approach, that can explore ever smaller spatial
(and higher energy) structure of the constituent matter. On the other hand,
the latter optical approach available only after the invention of laser in 1960
was to excite the atomic electronic structure and carries out spectroscopy of
the dynamics. The spectroscopy reveals structure and its dynamical characters
by palpating the atom by laser without pinpointed penetration into the core 
by the beam.
This also includes the selective excitation of certain characteristic
eigenmodes of the constituent matter structure. A good example may be the
method called Laser Induced Breakdown Spectroscopy (LIBS)~\cite{LIBS}.
When we study the vacuum itself, in stead of neutral atoms, this past
experience guides us the distinction of the two approaches the high momentum
vs. high field. As the main target
of this research, in contrast to the collider's Rutherford approach with high
momentum beam scattering, we are now introducing an alternative approach of the
method of {\it 'exciting and probing the texture of vacuum' } with high
amplitude of photons.

%
%
In order to probe such texture or polarization structures of the excited
vacuum, we are already treating vacuum as a medium to be studied, rather than
a given nothingness.
For example, we posit that the phase velocity of light is an
experimentally meaningful measurable quantity. Intense photons fields
(call it the pump laser or target laser) 
may provide nonlinearities even in vacuum under a
sufficient intensity (whose effects may be probed by another
probe laser). Such was predicted by Schwinger~\cite{Schwinger} in
the form of the generation of an electron-positron pair.
However, manifestations of nonlinearities and thus the possibility of nonlinear
spectroscopy of vacuum should emerge even sufficiently below the Schwinger
field threshold. (This may also correspond again to a similar phenomenon of
the atomic ionization by intense laser at the Keldysh field and the
emergence of atomic nonlinearities far below this Keldysh field.)
By this measurement we can investigate the dispersive and birefringence
characteristics of photons in the excited vacuum.

Let us consider the constituent's point of view,
since the phase velocity shift can also be understood
as a result of photon-photon interactions
~\cite{Toll, Narozhnyi, Ritus, Dittrich-Gies, Shore},
as we understood the refractive index in matter through the photon-atom
interaction at the level with atoms as the constituent.
Photon-photon interactions in vacuum are quantum processes depending on the
relevant frequency or mass scales of exchanged fields.
Thus, the dispersive nature must be discussed based on the relevant
frequency scale we introduce in experiments.
As is known from particle physics, in 100~GeV scale we expect
that the photon-photon interaction is caused by the heavy neutral boson 
exchange via higher order fermion loops to couple to photons~\cite{Higgs}.
As the frequency is lowered, lighter quark and electron loops may
cause the photon-photon interaction in the relevant quantum chromodynamics
(QCD) and QED mass scales of 100MeV and MeV.
Below these scales there is no known mass scale relevant for photon-photon
interactions.
As long as we use lasers with the eV energy scale with the intensity below
the Schwinger limit, therefore, the direct production of real constituent
particles is not possible. However, it is possible to discuss virtual 
vacuum polarizations. This was first introduced by the model through 
the Euler-Heisenberg effective action~\cite{EH} on the 
photon-photon interaction some 70 years ago. Thus far,
the real-photon-real-photon interaction has never been experimentally observed.
Therefore, when we embark on the study of probing vacuum by optical laser
fields, it is one of the first tasks to observe the QED interaction via the
measurement of the phase velocity shift under the intense laser field
due to this effect.

In addition to exploring the nonlinear QED the current method may introduce a
window through which we explore lighter mass scales well below MeV that
have been speculated or hinted from particle physics~\cite{PDG, ALPtheory} and
cosmology~\cite{DEreview}.
These are related yet unobserved fields of axions and
dark energy, for example. We note that the coupling of those light fields to
matter must be extremely small. Otherwise, higher energy experiments should
have discovered these fields already, since light particles can be copiously
produced. Here we recognize an advantage by introducing large amplitude
rather than high momentum to search for these light fields that weakly
couple to matter.
The relevant mass scales of the exchanged fields between photons
correspond to the force ranges. Therefore, the study of photon-photon
interactions in the low frequency scale below optical frequency may unveil
undiscovered semi-macroscopic forces as the basic building block enmeshed
in vacuum.

%
%
The focus of this paper is to discuss the experimental realization of a new
type for the semi-macroscopic scope of vacuum by studying the property of
nonlinearities of photons in vacuum. We shall discuss two examples of these.
First we suggest an experimental technique of the phase contrast 
Fourier imaging to
measure the phase velocity of photon under strong laser fields in
section~\ref{sec2} and \ref{sec3}.
This aims at a first verification of the nonlinear QED effect as pronounced by
Euler-Heisenberg and beyond.
Second we explore a realization of experiments to search for extremely light
fields or long ranged force via the resonance interaction, employing
quasi-parallel strong laser-laser interaction in section~\ref{sec4},
\ref{sec5} and \ref{sec6}.
In the conclusion section~\ref{sec7}, we summarize our approach with
high fields of lasers based on a wider perspective by comparing our 
methods with those in particle collider physics and cosmology.

\section{Probing nonlinear QED and beyond}\label{sec2}
Maxwell's equations suggest linear superposition of photons. On the other hand,
the quantum electrodynamics (QED) indicates a weak interaction of
photon-photon through the Feynmann's diagram of the box type, where a virtual
electron-positron pair loops and four real photons couple to
the loop at the four vertexes as external lines (see Fig.\ref{Fig6}(b)).
The photon-photon scattering cross section based on the box diagram is
calculated by \cite{KN,DT}.
The total elastic cross section in the center of mass system with the
photon energy of $\omega$ is expressed as
%
\beqa
\sigma_{\mbox{qed}} = \frac{973}{10125\pi}\alpha^2 {r_e}^2
\left(\frac{\hbar \omega}{m_e c^2}\right)^6,
\eeqa
%
where $\alpha=\frac{e^2}{\hbar c}$
is the fine structure constant,
$m_e$ is electron mass and
$r_e = \alpha\frac{\hbar c}{m_e c^2} \sim 2.8 \times 10^{-13}$~cm is
the classical electron radius. For photons of $\hbar \omega\sim1$~eV,
the cross section is $10^{-42}$~b. This is extremely small.
The smallness of this cross section arises from the electron-positron mass scale
of the four propagators in the loop of the box diagram or from
the short distance nature due to the corresponding Compton wavelength
of electron-positron pair.
We note that because of this smallness we have little 'noise', providing
a pristine experimental environment to search for something beyond QED.
In reverse if we do detect any signals in photon-photon interactions,
we are assured of something significant.
When one is interested in observing the events of QED real-photon-real-photon
scattering itself, the introduction of higher frequency photons as a probe
beam onto a high intense optical laser target would have a greater probability
due to the $(\hbar\omega)^6$ behavior than the optical-optical photon
interaction. In such an asymmetric colliding system, even elastic collisions
in the corresponding center-of-mass system (CMS)
can be detected as frequency shifted
scattered photons in the laboratory frame, depending on the collision geometry
with reasonably high statistics.
Furthermore, in such a setup, the non-perturbative nature of the
intense field is expected to be important. For example, the catalysis of
electron-positron pair production~\cite{Schuetzhold} due to a higher tunneling
probability from the Dirac sea may be tested,
even below the Schwinger field~\cite{Schwinger}.

In what follows we discuss on the measurement of the phase velocity shift,
where we focus on the optical-optical beam interaction.
The strong electromagnetic field may modify the dispersion relation
for photons. This effect is first discussed by Toll~\cite{Toll}.
If we detect the predicted velocity shift explained below,
it amounts to the verification of the nonlinear QED effects
in the perturbative regime.
In the low frequency collision it is sufficient to describe
the photon-photon interaction by the effective one-loop Lagrangian
\cite{EH, Weiscop, Schwinger}
%
\beqa\label{eq_EHL}
L_{1-loop} =
\frac{1}{360}\frac{\alpha^2}{m_e^4}
[4(F_{\mu\nu}F^{\mu\nu})^2+7(F_{\mu\nu}\tilde{F}^{\mu\nu})^2],
\eeqa
where $F_{\mu\nu} = \partial A_{\mu} / \partial x^{\nu} -
                    \partial A_{\nu} / \partial x^{\mu}$ is
the antisymmetric field strength tensor and its dual tensor
$\tilde{F}^{\mu\nu} = 1/2 \epsilon^{\mu\nu\lambda\sigma} F_{\lambda\sigma}$
with Levi-Civita symbol $\epsilon^{\mu\nu\lambda\sigma}$.
It is well known that the forward scattering amplitude $f$
with the dimension of length
is related to the refractive index $n$ via the Lorentz
relation~\cite{Goldberger}
%
\beqa
n(\omega) = 1 + \frac{2\pi}{(\omega/c)^2} N f(\omega),
\eeqa
%
where $N$ is the number density of scattering centers and
$f(\omega)$ is the forward scattering amplitude for light
of energy $\hbar\omega$.
The total cross section $\sigma$ is related to the absorptive part
of the forward scattering amplitude via the Optical Theorem, as follows
%
\beqa
\sigma = \frac{4\pi}{\omega/c} \mbox{Im} f(\omega).
\eeqa
%
If we use a strong coherent electromagnetic field with large $N$,
the real part of the forward scattering amplitude is expected to become large.
This is because the coherent addition of the scattering amplitudes over
large $N$ is expected.
Therefore, we are interested in the measurement of the refractive
index shift under the high electromagnetic field. It is not as effective
to focus laser to cause non-forward scattering processes.
The refractive index corresponds to
the inverse of phase velocity. The derivation of these quantities 
in the linearly polarized electromagnetic field target
(so called crossed-field configuration where electric field $\hat{E}$ and
magnetic field $\hat{B}$ are perpendicular with the same strength)
is originally studied in~\cite{Narozhnyi, Ritus} and further derived
from the generalized prescription based on the polarization tensor
applicable to arbitrary external fields in~\cite{Dittrich-Gies}.
This shows us
%
\begin{eqnarray}\label{eq_phsv}
v_{\parallel}/c = 1 - \frac{8}{45}\frac{\alpha^2}{{m_e}^4}\frac{\hbar^3}{c^5}\frac{z_k}{k^2},
\nonumber \\
v_{\perp}/c    = 1 - \frac{14}{45}\frac{\alpha^2}{{m_e}^4}\frac{\hbar^3}{c^5}\frac{z_k}{k^2},
\end{eqnarray}
%
where
$v_{\parallel}/c$ and $v_{\perp}/c$ are the phase velocities when the
combination of linear polarizations of the probe and target lasers is
either parallel and perpendicular, respectively. The quantity
$m^4_e c^5 /\hbar^3 \sim 1.42\times10^{24}$J/m${}^3$ is the Compton
energy density of an electron, $k$ is the wave number of the probe
electromagnetic field with the unit vector of $\hat{k}$. The Lorentz
invariant quantity $z_k$ is defined as
%
\begin{eqnarray}\label{eq_zk}
z_k = (k_{\alpha}F^{\alpha\kappa})(k_{\beta}F^{\beta}_{\kappa}),
\end{eqnarray}
%
and the relation to the energy density $\epsilon^2$ in the crossed field
condition is
%
\begin{eqnarray}\label{eq_zk_epsilon}
\frac{z_k}{k^2} = \epsilon^2 (1+(\hat{k}\cdot\hat{n}))^2,
\end{eqnarray}
%
with $\epsilon = E = cB$ and $\hat{n} = \hat{B} \times \hat{E}$.
Thus the second terms in (\ref{eq_phsv}) show the deviation of
the phase velocities of light $v_{\parallel}$ and $v_{\perp}$
are proportional to the field energy density normalized to the
Compton energy density of an electron.

The shift of the refractive index from that of the normal vacuum is
on the order of $10^{-11}$ for the energy density $\epsilon^2$ of 1J/$\mu$m${}^3$.
The refractive medium has the polarization
dependence, $i.e.$ the birefringence nature. The difference in $v_{\parallel}$
and $v_{\perp}$ in (\ref{eq_phsv}) results from
the first and second terms in the bracket
of the effective one loop Lagrangian in (\ref{eq_EHL}).
The UV limit ($\omega\rightarrow\infty$)
of the dispersion and birefringence under a constant
electromagnetic field may be evaluated via the Kramers-Kronig
dispersion relation, as discussed in~\cite{Shore}.
The phase velocity in both UV and IR limits is expected to be subluminal
($v_{phase} < c$) under the QED field ~\cite{Shore, Dittrich-Gies}.
The UV limit of the phase velocity is supposed
to govern causality. Therefore, it can be a fundamental test of a variety
of effective field theories in the IR limit whether the phase velocity in the
UV limit extrapolated from that of IR is superluminal
($v_{phase}(\infty) > c$) or not.
Thus far the dispersion relation from IR to UV is theoretically known only
in the QED field~\cite{Shore}. However, there has never been data even
in IR frequencies to date. It is important, therefore, for experiments to
quantitatively verify or disprove the QED prediction.
We note that the measurement of the refractive index in higher frequency
may be sensitive to the part of the anomalous dispersion as discussed in
~\cite{Heinzl} and also the measurement of the electron-positron
pair creation~\cite{Schuetzhold, Dunne, Baier, Narozhny2}
in strong electromagnetic field may be directly sensitive to
the absorption or imaginary part.
The Kramers-Kronig relation connects between the real and imaginary parts
of the forward scattering amplitude or the refractive index.
Therefore, the systematic measurements of real and imaginary parts over
wide frequency range may provide a test ground of QED and
the Kramers-Kronig relation itself when it is applied to vacuum.

Suppose then the detected dispersion and birefringence quantitatively deviate
from the expectation of QED. This should indicate that undiscovered fields
may be mediating photons beyond QED.
Scalar and pseudoscalar types of fields may appear via the first and second
products in brackets of (\ref{eq_EHL}), respectively as we discuss
section~\ref{sec4} in detail.
In addition if masses of scalar and/or pseudoscalar fields are light,
this long distance nature may enhance the coherence nature and thereby
the amplitude of the forward scattering
compared to the lowest order QED diagram.
%
Therefore, the measurement of the birefringence and the comparison to
the expectation from the nonlinear QED effects may be a general
probe to investigate the fundamental nature of vacuum.

\section{Phase contrast Fourier imaging on the focal plane}\label{sec3}
We now consider an experiment where we create a
high intensity spot by focusing a laser in vacuum and
we probe its refractive index shift by a second laser.
We call the first laser as the target laser,
while the second as the probe laser hereon.
The key issue is to detect an extremely small refractive index change
resulting from the photon-photon interaction
between the target and probe lasers.
The conventional way that was already performed~\cite{PVLAS} and
proposed~\cite{LaserLaser, LaserDiffraction} is based on the measurement of
the ellipsoid caused by birefringence and that of the rotation angle
of a linearly polarized probe laser by  making it
propagate for a long distance under a weak magnetic field\cite{PVLAS} or
electromagnetic field~\cite{LaserLaser, LaserDiffraction}.
This method has an advantage to enhance the phase shift
by a long optical path without introducing costly strong target
electromagnetic fields.
In the case of the constant magnetic field on the order of 1T,
one encounters the limit of physical detectability sensitive
to the QED nonlinear effect. In the case of strong electromagnetic field,
one runs into the damage limit to store the strong target field within
a cavity over a long time.
On the other hand, if we can directly utilize the local nature of vacuum
by tightly focusing an intense laser pulse and measuring the lensing effect
of vacuum on the pulse-by-pulse basis,
there is no physical limit in increasing the intensity of the laser pulse
until vacuum itself breaks down. In order to increase the shift in the
refractive index corresponding to the inverse of phase velocities in
(\ref{eq_phsv}), $i.e.$ the intensity of the target laser field
as expected through (\ref{eq_zk}) and (\ref{eq_zk_epsilon}),
it is necessary to use a focused
laser pulse by confining a large laser energy into a small spacetime volume.
This causes a locally different refractive index along the
trajectory of the target laser pulse in vacuum. A variation of
the refractive index arises over the high intense
part and the rest of vacuum. If the probe laser penetrates into both parts
simultaneously, the corresponding phase contrast should be embedded in the
transverse profile of the same probe laser. Our proposal is to directly measure
the phase contrast and to determine the absolute value of the refractive
index change by controlling the combination of polarizations of the probe
and target laser pulses. This should result in the birefringence
as expected in (\ref{eq_phsv}).

We need to detect the extremely small shift of phase velocity by the
target-probe interaction. For this we also need an intense probe laser
in order to enhance visibility.
However, if we utilize the conventional interferometer techniques providing a
homogeneous phase contrast over the probe laser profile, such small refractive
index changes are hard to detect. This is because the resulting intensity
modulation is always on top of the huge pedestal intensity and the contrast of
the modulation to the pedestal is extremely small.
Any photo device cannot be sensitive to a small number of
photons spatially modulating under the pedestal intensity
beyond 1J ($\sim 10^{19}$ visible photons) due to the limit of the
dynamic range on the photon intensity measurable by a camera pixel
without causing saturation on the intensity measurement.
On the other hand, broadening the dynamic range by lowering the gain of
the electric amplification of photo-electron degrades the sensitivity
to the small number of the spatially modulating photons or the sensitivity to
the small phase shift.
Therefore, we need to invent a method that can spatially separate the weakly
modulating characteristic intensity pattern from the strong pedestal.

\begin{figure}
\includegraphics[width=1.0\linewidth]{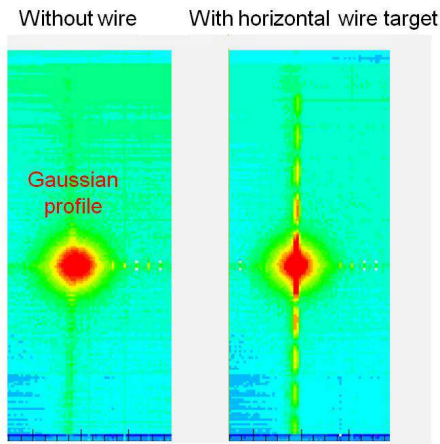}
\caption{
Diffraction images at the far distance
from the thin wire target when the Gaussian laser is shot on the wire.
The left figure is the case there is no wire.
The right figure is the case when the thin wire target is horizontally arranged.
}
\label{Fig1}
\end{figure}

In order to overcome this difficulty,
we suggest utilizing the inhomogeneous phase contrast Fourier imaging
on the focal plane by focusing the probe laser via a conceptual lens.
In this the physically embedded phase contrast on the transverse profile
of the probe laser amplitude is Fourier transformed on the focal plane.
The intensity pattern on the focal plane has a preferable
feature that the characteristic phase boundary makes the intensity profile
expand outer regions far from the focal point, whereas a Gaussian laser
with a homogeneous phase converges into a small focal point with the
smallest beam waist. It is instructive to illustrate the characteristic nature
of the diffraction pattern from a wire-like target shape as shown in
Fig.\ref{Fig1}. Here the intensity pattern at a far distance known as the
Fraunhofer diffraction limit is shown in the case when a Gaussian laser beam
is shot on the thin wire target.
The intensity pattern at a far distance can be understood as the Fourier
transform of the wire shape approximated as a rectangular of
$2\mu \times 2\nu$. It is well known that a lens component by itself also
has the equivalent effect to produce a diffraction pattern corresponding to
the exact Fourier transformed image of the original shape of the
refractive medium on the focal plane at a finite distance
(see \cite{SIEGMAN} for instance).
In order to understand the diffraction image, we may refer to
Babinet's principle which states that the diffraction pattern from an
opaque wire is identical to that from a slit of the same size and shape.
The Fourier transform of such a rectangular slit is expressed as
%
\begin{eqnarray}
\left(\frac{\sin(\mu\omega_x)}{\mu\omega_x}\right)^2
\left(\frac{\sin(\nu\omega_y)}{\nu\omega_y}\right)^2,
\end{eqnarray}
%
where $\omega_x=\frac{2\pi}{\lambda f}x$ and $\omega_y=\frac{2\pi}{\lambda f}y$
are the spatial frequencies for the given position $(x,y)$ on the focal plane
with the focal length $f$ and the wavelength $\lambda$.
In the case of the slit with $\mu\gg\nu$,
the rectangular profile on the focal plane becomes orthogonally rotated
thinner line shape with oscillations included in the line
(see Fig.\ref{Fig1}~(right)). This is because the narrower the slit size is,
the smaller the spatial frequency in that direction becomes.
On the other hand, the shape of Gaussian distribution
without the wire or slit is unchanged,
because the Fourier transform of the Gaussian distribution corresponds
to the Gaussian distribution (see Fig.\ref{Fig1}~(left)).
This is the key feature that drastically improves the detectability to
the small phase shift by sampling only outer parts far from the Gaussian part.
This may also be interpreted as the counter-concept to the conventional
spatial filter, where outer parts are eliminated to maintain a smooth
phase on the transverse profile of the Gaussian distribution.

\begin{figure}
\includegraphics[width=1.0\linewidth]{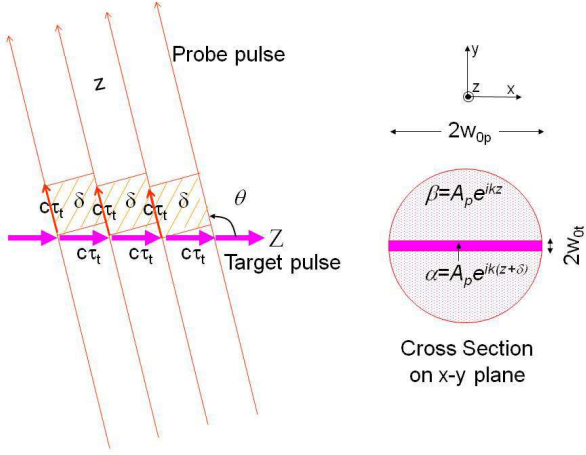}
\caption{
Geometry of the embedded phase contrast in the probe pulse.
}
\label{Fig2}
\end{figure}

Given the intuitive picture above,
a quantitative formulation of our proposed method is presented as follows.
First, let us define the geometry of the laser intersection,
as shown in Fig.\ref{Fig2}~(left),
where the tightly focused target pulse with time duration $\tau_t$
propagates along the $Z$-axis and the probe pulse with the larger profile
and longer time duration $\tau_p$ propagates along the $z$-axis tilted
by $\theta$ from the $Z$-axis. In this figure the wavefronts of the probe
pulse are drawn successively with time step $\tau_t$ under the condition
$\tau_p > \tau_t$. Since the constant phase shift is embedded only during
$\tau_t$, the optical length with the constant refractive index shift along
the $z$-axis perpendicular to the wavefront is limited to $c\tau_t$.
This is independent of $\theta$. In this geometry after the
penetration of the probe laser pulse, the profile of the probe laser on
the $x-y$ plane contains a trajectory with the constant phase shift $\delta$
along the projection of the path of the target laser on the probe wavefront,
as shown in Fig.\ref{Fig2}~(right).

In order to discuss the amount of the phase shift, we need a concrete
geometry at the diffraction limit of both target and probe lasers.
Let us briefly review the laser profile at the diffraction limit.
The Gaussian profile is a basic constraint in typical laser fields,
where the aperture of a lasing material has a finite
size in the transverse area. The solution of the electromagnetic field
propagation in vacuum with the Gaussian profile on the transverse plane
with respect to the propagation direction $z$ is well-known\cite{Yariv}.
The electric field component in spatial coordinates $(x,y,z)$ is expressed as
%
\beqa\label{eq_Gauss}
E(x,y,z) \propto 
\qquad \qquad \qquad \qquad \qquad \qquad \qquad \qquad \qquad \nnb\\
\frac{w_0}{w(z)}\exp
\left\{
-i[kz-\eta(z)] - r^2 \left( \frac{1}{{w(z)}^2}+\frac{ik}{2R(z)} \right)
\right\},
\eeqa
%
where $k=2\pi/\lambda$, $r=\sqrt{x^2+y^2}$, $w_0$ is the minimum waist
which cannot be smaller than $\lambda$ due to the diffraction limit, and
other definitions are as follows:
%
\beqa\label{eq_wz}
{w(z)}^2 = {w_0}^2
\left(
1+\frac{z^2}{{z_R}^2}
\right),
\eeqa
\beqa\label{eq_Rz}
R = z
\left(
1+\frac{{z_R}^2}{z^2}
\right),
\eeqa
\beqa\label{eq_etaz}
\eta(z) = \tan^{-1}
\left(
1+\frac{z}{z_R}
\right),
\eeqa
\beqa\label{eq_z0}
z_R \equiv \frac{\pi{w_0}^2}{\lambda}.
\eeqa
%

Let us express the phase shift $\delta$ in the limit
$z=c\tau_t < z_R$, where we assume an almost flat wavefront
as indicated by (\ref{eq_wz}) and (\ref{eq_Rz})
in the vicinity of the diffraction limit
%
\beqa\label{eq_delta}
\delta = \frac{2\pi}{\lambda_p}
\delta n_{qed} c\tau_t \varphi_t(x_p,y_p),
\eeqa
where the subscripts $p$ and $t$ denote the {\it probe} and {\it target}
quantities, respectively, $\delta n$ is
the refractive index shift, $c\tau_t$ is the path length with
effectively constant phase shift $\delta$ over the crossing time $\tau_t$ and
$\varphi_t(x_p,y_p)$ is a weight function to reflect the path length difference
depending on the incident position with respect to the target profile
expressed as a function of position $(x_p, y_p)$ on the transverse
plane of the probe laser.
Based on (\ref{eq_phsv}), (\ref{eq_zk}) and (\ref{eq_zk_epsilon})
we parametrize the refractive index shift as
\beqa\label{eq_delta_para}
\delta n_{qed} = \zeta N_0 (1-\cos\theta) \frac{E_t}{\pi {w^2_0}_t c\tau_t},
\eeqa
where
$\zeta$ is 4 and 7 for the polarization combinations $\parallel$ and $\perp$,
respectively,
$N_0$ is the coefficient to convert from energy density to
refractive index shift defined as
$N_0 \equiv \frac{2}{45}\frac{\alpha^2\hbar^3}{m^4_e c^5}
= 1.67\times 10^{-12}  [\mu\mbox{m}^3/\mbox{J}]$,
$\theta$ is the incident angle of the probe pulse with respect to
the propagation direction of the target pulse as depicted in Fig.\ref{Fig2},
$E_t$ is the energy of the target pulse in [J], and
$\pi {w^2_0}_t c\tau_t$ is the volume in [$\mu$m${}^3$]
for the given target profile
with the minimum waist ${w^2_0}_{t}$ from (\ref{eq_wz}).
By substituting (\ref{eq_delta_para}) into
(\ref{eq_delta}) we obtain the experimentally convenient
expression for $\delta$
\beqa\label{eq_delta_exp}
\delta \sim
\frac{2\pi}{\lambda_p} \zeta N_0 (1-\cos\theta) \frac{E_t}{\pi {w^2_0}_{t}}
\varphi_t(x_p, y_p).\eeqa
%

We are interested in an application in the limit
${w_0}_t<{w_0}_p$ for the width and $c\tau_t<{w_0}_p$ for the depth of
the embedded phase shape. We then take the approximation
$\varphi_t(x_p,y_p) \sim 1$ to simplify the following argument
(if necessary, we may restore the target profile $\varphi_t(x_p,y_p)$
based on the precise profile of the target laser reflecting actual experimental
setups). In this limit we approximate that the target profile has
a rectangular shape with the size of $2\mu \times 2\nu$ inside which a
constant phase shift is assigned, where the effective slit sizes are defined by
the transverse sizes of focused laser beams through the relation
$\mu \sim {w_0}_p$ and $\nu \sim {w_0}_t$.
We note that the optical length $c\tau_t$ with phase shift $\delta$
of the probe laser is eventually canceled out based on (\ref{eq_delta_exp}).
This indicates that we have many
choices on $\tau_t$ as long as the conditions $c\tau_t < {z_R}_t$
and $\tau_t < \tau_p \sim {z_R}_t/c$ are satisfied.

We introduce the window functions $rec$ and $\overline{rec}$ as follows:
%
\begin{eqnarray}\label{eq_slit}
rec(\mu, \nu) = \left\{
\begin{array}{ll}
1 & \quad \mbox{for $|x|\le\mu \cap |y|\le\nu$} \\
0 & \quad \mbox{for $|x|>\mu \cup |y|>\nu$}
\end{array}
\right\},
\nnb\\
\overline{rec}(\mu, \nu) = \left\{
\begin{array}{ll}
0 & \quad \mbox{for $|x|\le\mu \cap |y|\le\nu$} \\
1 & \quad \mbox{for $|x|>\mu \cup |y|>\nu$}
\end{array}
\right\}.
\end{eqnarray}
%
This window provides a unit region of a constant phase which is
employed for arbitrary phase maps composed of a collection of
the unit window cells. If we determine the phase on the pixel-by-pixel basis
for a given camera device, the rectangular shape as a minimum unit cell should
be a natural choice.

The intensity distribution at the focal plane is determined
as a function of the peak amplitude of the probe pulse ${A_0}_p$,
the wavelength $\lambda_p$ and the focal length $f_p$ for a given
Gaussian beam profile of ${A_0}_p e^{-a(x_0^2+y_0^2)}$
as the probe laser. The linearly synthesized amplitude at $z$
after crossing with the target laser pulse is expressed as
%
\begin{equation}\label{eq_planewave}
\psi(x_0, y_0) = \alpha rec(\mu,\nu)e^{-a(x_0^2+y_0^2)} + \beta \bar{rec}(\mu,\nu)e^{-a(x_0^2+y_0^2)},
\end{equation}
%
where $\alpha$ and $\beta$ are the plane waves at the
point $z$ after the penetration of the probe laser.
The function $\alpha$ with the local phase shift
$\delta$ caused by the local refractive index shift and $\beta$
without phase shift are defined as
%
\begin{eqnarray}\label{EqPlane}
\alpha &=& {A_0}_p e^{i(kz+\delta)}, \nonumber \\
\beta  &=& {A_0}_p e^{ikz}.
\end{eqnarray}
%
The Fourier transform $F$ of the synthesized amplitude $\psi$ on
the focal plane $(x,y)$ at $z$ is expressed as
%
\begin{eqnarray}\label{EqF}
F\{\psi(x_0, y_0)\} = 
\qquad \qquad \qquad \qquad \qquad \qquad \qquad \qquad \nnb\\
\alpha F\{rec(\mu,\nu)e^{-a(x_0^2+y_0^2)}\} +
\beta F\{\bar{rec}(\mu,\nu)e^{-a(x_0^2+y_0^2)}\}
\nnb\\
= (\alpha-\beta) \int^{\mu}_{-\mu}\!\!\int^{\nu}_{-\nu}\!\!dx_0dy_0
e^{-a(x_0^2+y_0^2)} e^{-i(\omega_x x_0 + \omega_y y_0)} +
\nnb\\
\beta \int^{\infty}_{-\infty}\!\!\int^{\infty}_{-\infty}\!\!dx_0dy_0
e^{-a(x_0^2+y_0^2)} e^{-i(\omega_y x_0 + \omega_y y_0)},\hspace{0.3cm} 
\end{eqnarray}
%
where we define $(\omega_x,\omega_y)\equiv(\frac{2\pi}{f_p\lambda_p}x,
\frac{2\pi}{f_p\lambda_p}y)$. We introduce coefficient $C_{sig}$
for the first term in (\ref{EqF})
containing the information on how much the phase shift,
namely {\it signal}, is localized resulted in the photon-photon interaction
defined as
%
\begin{eqnarray}\label{EqCsig}
C_{sig}(\omega_x, \omega_y) \equiv
 \qquad \qquad \qquad \qquad \qquad \qquad \qquad \nnb\\
 \int^{\mu}_{-\mu}\!\!dx_0 e^{-a x_0^2} \cos(\omega_x x_0)
 \int^{\nu}_{-\nu}\!\!dy_0 e^{-a y_0^2} \cos(\omega_y y_0).
\end{eqnarray}
%
We also define coefficient $C_{bkg}$ for the second term of (\ref{EqF})
which corresponds to the background pedestal Gaussian part as
%
\begin{eqnarray}\label{EqCbkg}
C_{bkg}(\omega_x, \omega_y) \equiv
\frac{\pi}{a} e^{-\frac{(\omega_x^2+\omega_y^2)}{4a}}.
\end{eqnarray}
%
Therefore, the Fourier transform becomes
%
\begin{eqnarray}\label{EqFfinal}
F\{\psi(x_0, y_0)\} =
(\alpha-\beta) C_{sig}(\omega_x, \omega_y)
\nnb\\
+ \beta C_{bkg}(\omega_x, \omega_y).
\end{eqnarray}
%
By substituting (\ref{EqPlane}), (\ref{EqCsig}) and (\ref{EqCbkg})
into (\ref{EqFfinal}), the intensity pattern at the focal plane
is expressed as
%
\begin{eqnarray}\label{eq_focalint}
|\psi(\omega_{x},\omega_{y})|^2=
\quad \qquad \qquad \qquad \quad
\quad \qquad \qquad \qquad \nnb\\
(\frac{{A_0}_p}{f_p\lambda_p})^2
\{ 2C_{sig}(C_{sig}-C_{bkg})(1-\cos\delta)
+C^2_{bkg} \}.
\end{eqnarray}
%
Equation (\ref{eq_focalint}) indicates that this method eventually corresponds
to an interferometer via the cross term of
$2C_{sig}(C_{sig}-C_{bkg})(1-\cos\delta)$.
This interferometer differs from the conventional one in
that the modulating part due to the phase shift $\delta$ can be spatially
separated from the confined strong Gaussian part $C^2_{bkg}$ due to the
nature of $C_{sig}$ part.
Therefore, in principle, our method of sampling only peripheral
intensity modulations caused by non-zero $\delta$,
provides a high signal-to-pedestal ratio
circumventing the most intense focal spot.
This is demonstrated in \ref{subsecB}. See Fig.\ref{Fig4} and \ref{Fig5}.

We note that this method is found to be similar to but distinct from
the idea in \cite{NaturePhotonics},
where two intense target laser pulses are treated as a matterless double
slit and the interference between spherical waves from these slits
is discussed as a signature of the photon-photon interaction.
In their proposal the physical diffraction is caused by
laser-laser interaction itself.
Our method rather lets the target laser cause the refractive
phase shift in the probe laser as indicated in Fig.\ref{Fig3}.
This phase shift is embedded in a refracted nearly plane wave
in the forward direction of the probe laser
as explicitly formulated in (\ref{eq_planewave}) and (\ref{EqPlane}).
We then set a lens to the right of interaction between the target and
probe lasers as shown in Fig.\ref{Fig3}.
The diffraction or Fourier transform in our method is incurred
by the added phase of the lens component
and the spherical wave propagation from the lens to the focal plane.
The advantage of our method is the enhanced detectability of small phase shift
on the pulse-by-pulse basis, as is demonstrated in section \ref{subsecB}
due to more efficient collection of photons by the lens effect
with the simpler target geometry.
On the other hand, a disadvantage is the deviation from the ideal lens phase
after embedding the physical phase on the probe laser. How to correct
this kind of background phase fluctuations is discussed in the following
subsections.

We also note that the Fourier image on the focal plane provides
only the absolute value of phase shift $\delta$.
Because of $1-\cos\delta \sim \delta^2/2$
with $\delta \ll 1$ in (\ref{eq_focalint}),
the proposed method has no sensitivity to the sign of phase shift in the
case of almost homogeneous background phase.
On the other hand, if a special offset phase $\pi/2$ is
locally added along the path of the target laser before the final
focusing to have the Fourier image, such setup
may recover the sensitivity to the sign of phase shift,
because $1-\cos(\delta+\pi/2)$ in (\ref{eq_focalint}) becomes $\sim 1+\delta$.
From physical point of view, it is important to discuss whether
phase shift is positive or negative, since it reflects the dynamics of
local interaction.
Furthermore, this has a definite advantage to enhance the signal
due to the proportionality to $\delta$ compared to $\delta^2$
for the extremely small $\delta$.
Therefore, the implementation of such local offset phase on the probe laser
in advance should be a part of experimental task.

\subsection{Template analysis for local phase reconstruction}\label{subsecA}
In actual experimental situations it is unavoidable to
contain local phase fluctuations inside the probe pulse
on the pulse-by-pulse basis even in the absence of the physical signal.
Deviations from the ideal Gaussian mode\cite{NonGaussian} and
local phase fluctuations due to optical elements in the path of the probe laser
may be sources of local phase fluctuations.
Compared to the physical phase shift $\delta$ from the nonlinear QED effect,
these fluctuations are expected to be much larger. However, if the local phase
map $\phi(X)$ on each probe pulse is {\it a priori} measured
as a function of position $X \equiv (x_0, y_0)$
on the transverse plane of the probe pulse, we may be able to correct the effect
from the background fluctuations.
In the next subsection \ref{subsecB} we discuss how to measure
the discrete phase map $\phi_i \equiv \phi(X_i)$ on the pulse-by-pulse basis
in detail, where $i$ denotes a discrete position on the transverse plane
at which $\delta$ is embedded. Here we focus on how to determine phase shift
$\delta$ from the combination of experimental data on the phase maps
under such background fluctuations on the pulse-by-pulse basis.

Let us generalize the discussion of (\ref{EqF}). The integral range
in (\ref{eq_slit}) included in the first of (\ref{EqF})
can take any shape and size in general. We replace the rectangular
region $rec$ with the region $R_i \equiv R(X_i)$
where a constant phase is mapped within
the region for the given discrete position $X_i$.
By denoting the spatial frequency as
$W=(\omega_x,\omega_y) = (2\pi x/(f_p\lambda_p), 2\pi y/(f_p\lambda_p))$
for the position on the focal plane $(x, y)$ with the integral kernel
$f(W, X) \equiv e^{-a(x^2_0+y^2_0)}e^{-i(\omega_x x_0 + \omega_y y_0)}$,
the Fourier transform including the local phase fluctuations $\phi_i$ is
expressed as
%
\begin{eqnarray}\label{eq_psiW}
\psi(W; \phi) &=& F\{\psi(X; \phi)\} \nnb\\
&=& \sum^{N_X}_i \{\alpha(\phi_i)-\beta\} \int_{R_i} dX f(X,W) \nnb\\
&=& \sum^{N_X}_i \{\alpha(\phi_i)-\beta\} I_i(W) + \beta I_{\infty}(W),
\end{eqnarray}
%
where $N_X$ is the number of regions on the transverse plane at $z$,
$\alpha(\phi_i) = A_0 e^{-i(kz + \phi_i)}$, $\beta = A_0 e^{-ikz}$,
$ I_i(W) = \int_{R_i} dX f(X,W)$, and
$I_{\infty}(W) = \int^{\infty}_{-\infty} dX f(X,W)$.
We note that this expression corresponds to the regional cut and paste
of a Gaussian laser amplitude; {\it i.e.},
cutting a region with a phase determined from
$\beta$ at $z$ and paste the same region by adding $\phi_i$ in
$\alpha(\phi_i)$.

Given $\phi(X_i)$ on the pulse-by-pulse basis,
we can numerically calculate the
real and imaginary parts of $\psi(W;\phi)$ as follows;
\begin{eqnarray}
\mbox{Re} \psi(W;\phi) =
\sum^{N_X}_i\{\cos(kz+\phi_i)-\cos(kz)\} I_i(W) \nnb\\
+ \cos(kz)I_{\infty}(W), \qquad \qquad \qquad
\nnb\\
\mbox{Im} \psi(W;\phi) =
\sum^{N_X}_i\{\sin(kz+\phi_i)-\sin(kz)\} I_i(W) \nnb\\
+ \sin(kz)I_{\infty}(W). \qquad  \qquad  \qquad 
\end{eqnarray}
The estimated background intensity pattern $I_{bkg}(\phi)$
on the focal plane with the phase fluctuations $\phi$
without physical phase is given by
\begin{eqnarray}\label{EqIphi}
I_{bkg}(W;\phi) = \{\mbox{Re} \psi(W;\phi) \}^2 + \{\mbox{Im} \psi(W;\phi) \}^2.
\end{eqnarray}

We now embed a template of the local physical phase shift
$\delta_i \equiv \delta(X_i)$ as well. The phase shift $\delta_i$ can be
evaluated from geometry of the energy density profile, {\it i.e.}
the refractive index distribution inside the intense target laser field.
This can be fixed from the experimental design on the focal spot
{\it a priori}. We can monitor if the center of the spot is surely stable
and further correct its deviation from the fixed geometry of the target laser.
Given $\delta_i$, we have to only replace the phase
by $\phi_i \rightarrow \phi_i + \kappa \delta_i$ with a constant
parameter $\kappa$ as follows
%
\begin{eqnarray}\label{EqIphidelta}
I_{bkg+sig}(W;\phi+\kappa\delta) =
\qquad \qquad \qquad \qquad \qquad \nnb\\
\{\mbox{Re} \psi(W;\phi+\kappa\delta) \}^2 + \{\mbox{Im} \psi(W;\phi+\kappa\delta) \}^2,
\end{eqnarray}
%
where $bkg+sig$ refers to the fact that the physical phase is embedded in the
background phase.

Given the measured intensity pattern $I_{meas}$ on the focal plane
per probe pulse focusing,
we define $\chi^2$ with (\ref{EqIphidelta}) as a function of $\kappa$
%
\begin{eqnarray}\label{eq_chi2}
\chi^2(\kappa) \equiv 
\qquad \qquad \qquad \qquad
\qquad \qquad \qquad \qquad \qquad \nnb\\
\frac{1}{N_W-1} \sum_j^{N_W}
\frac{|I_{meas}(W_j)-I_{bkg+sig}
(W_j;\phi+\kappa\delta)|^2}{I_{meas}(W_j)+I_{bkg+sig}(W_j;\phi+\kappa\delta)},
\end{eqnarray}
%
where $N_W$ is the number of sampling points on the focal plane and
$j$ runs through all discrete positions on the focal plane except
for the most intense region around the focal point.
We then determine the parameter $\kappa$ by minimizing the $\chi^2$
on the pulse-by-pulse basis within the acceptable accuracy.
The reason why we have to determine $\kappa$ on the Fourier image
on the focal plane is due to the experimental constraint that
we cannot sample the most intense part at the focal point.
Because of the information loss at the focal point corresponding
to the lower spatial frequency part,
we cannot reconstruct the original phases of the probe laser
before focusing by the inverse Fourier transform for the lens effect.

We note that determining $\kappa$ corresponds to the measurement
of the absolute value of phase shift $\delta$.
As we have discussed in the last paragraph of section \ref{sec3},
if the offset phase $\pi/2$ is locally added to the known physical template
before the focus by the lens, it provides
the sensitivity to the sign of phase shift.
This also has the advantage that the intensity change of the physical
signal on the characteristic Fourier image is drastically enhanced
due to the proportionality to $\delta$ compared to $\delta^2$.
Such local phase may be implemented on the probe laser
by utilizing the photorefractive crystal. The photorefractive
effect\cite{Yariv} causes a static local refractive index change
by supplying an external electric field onto the crystal.
If we embed a rectangular shape with different refractive
index on the surface of the photorefractive crystal by mimicking
the shape of the target laser trajectory in vacuum, we may be able
to implement the local phase offset $\pi/2$ onto the physical template
in advance by making the probe laser propagate through the crystal
before the final focusing by the lens. If this is the case, we can use
a much weaker probe laser. In following subsections, however, we consider the
case where such special offset phase is not implemented; instead,
a sufficiently intense probe laser as well as the target laser
is available in an experiment.
\subsection{Suggested setup and the Fourier image}\label{subsecB}
\begin{figure}
\includegraphics[width=1.0\linewidth]{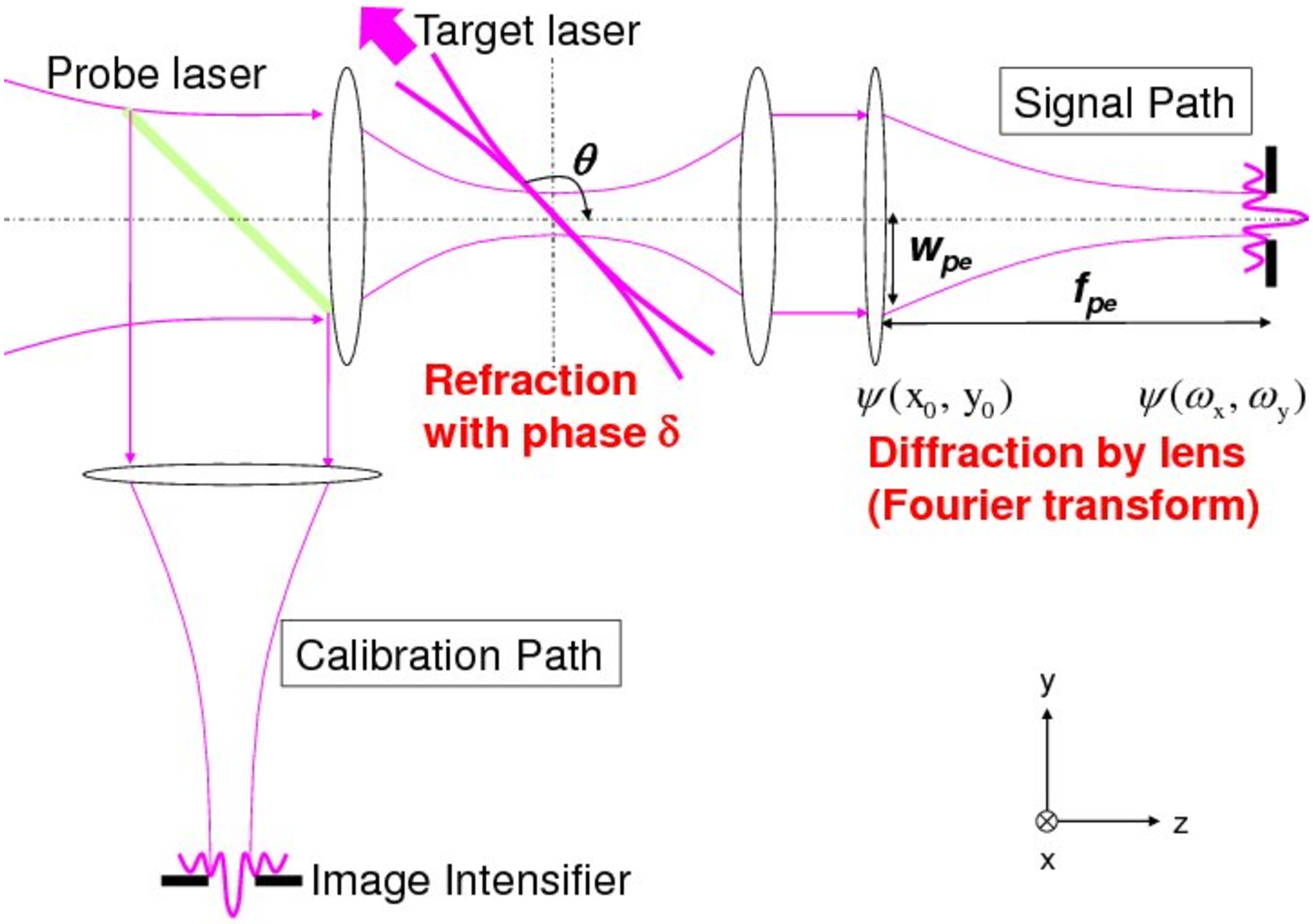}
\caption{
Conceptual experimental setup for the suggested phase contrast Fourier imaging.
At the crossing point between the probe and target lasers,
the target laser causes the shift in the index of refraction,
which amounts to the refractive phase shift $\delta$
embedded in the probe laser as explained in Fig.\ref{Fig2}.
After the expansion from the crossing point,
this probe laser goes through the lens and the signal path on the right.
The diffraction or Fourier transform is incurred
by the added phase of the lens component
and the spherical wave propagation from the lens to the focal plane
as demonstrated in Fig.\ref{Fig1} and later in Fig.\ref{Fig4} and \ref{Fig5}.
}
\label{Fig3}
\end{figure}

Figure \ref{Fig3} illustrates the conceptual experimental
setup for the phase contrast Fourier imaging on the focal plane.
The setup consists of two parts; the signal path (SP) and the calibration
path (CP). The part SP is to perform the phase contrast Fourier imaging to
measure a physically embedded phase shift
by the probe-target laser interaction.
Combined with SP, CP is to provide phase maps of static optical
components $\phi_{opt}(X_i)$ and the phase map $\phi_{pls}(X_i)$ of
probe laser brought on by instabilities in the upstream laser systems
on the pulse-by-pulse basis, as we have discussed in \ref{subsecA}.

Within SP, both the target and probe laser beams are focused with different
waist sizes ${w_t}_0$ and ${w_p}_0$
with the $F$-number, $F_t=f_t/d$ and $F_p=f_p/d$
for the common incident beam diameter $d$, respectively.
The lens size must be larger than $d$.
They are crossed each other at the minimum waist point where
the wave fronts in both beams are close to flat in the case of the ideal
Gaussian profile beam with $R=\infty$ in (\ref{eq_Rz}).
We assume that the target waist ${w_0}_t$ is smaller than
the probe waist ${w_0}_p$ which embeds a phase contrast
within the amplitude on the transverse profile of the probe laser.

The crossing angle $\theta$ has no impact on the optical length
with the shift of the refractive index. However,
it is important to control the amount of the refractive index shift,
depending on the relevant dynamics for photon-photon interactions.
In the case of QED the setting $\theta \sim \pi$ has the
advantage to increase the photon-photon center of mass energy,
as discussed in (\ref{eq_delta_exp}).
We design the target geometry such that the Rayleigh length
of the target pulse ${z_R}_t$ is approximated as $\mu$ in (\ref{eq_slit})
and the target beam waist ${w_t}_0$ is approximated as $\nu$.
Accordingly the beam waist of the probe pulse ${w_p}_0$ should coincide
with ${z_R}_t$ within the probe duration time $\tau_p = {z_R}_t/c$.

After embedding the physical phase shift onto the probe pulse at the minimum
waists of both beams, the probe laser is expanded by expansion factor $R_e$
to get the probe beam waist of ${w_p}_e$.
The expanded probe laser is refocused with the focal length ${f_p}_e$.
It is important to notice that the amount of the embedded phase contrast
is independent of the rate of magnification of the transverse
profile of the probe laser. This implies that we have degrees of freedom to
design the final focus to produce a Fourier image with
the sufficient number of photons per camera pixel
that enables the pulse-by-pulse analysis, as we see below.

\begin{figure}
\includegraphics[width=1.0\linewidth]{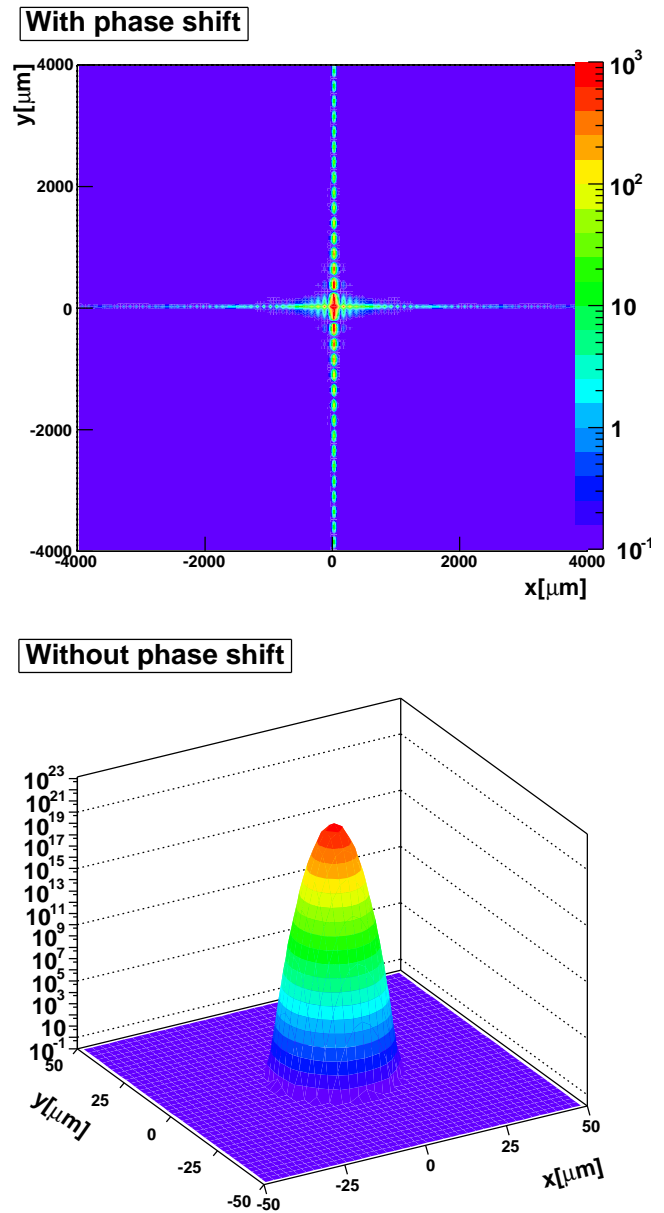}
\caption{
Intensity profile in unit of the number of photons per single shot focusing
on the focal plane by sampling the values with $50\mu$m steps
in both x and y axises.
The relevant laser parameters are summarized in Tab.\ref{Tab1}.
The top figure is the intensity value per point with physical phase shift
based on (\ref{eq_focalint}). The color map is normalized to $10^3$
by truncating the intensity at the focal point to zoom the signal part in the
color map. The bottom plot shows the same quantity without physical phase shift
in 3-D to show how much the ideal pedestal Gaussian intensity profile is
localized quantitatively in the $x-y$ plane without the truncation of
the intensity at the focal point.
}
\label{Fig4}
\end{figure}

\begin{figure}
\includegraphics[width=1.0\linewidth]{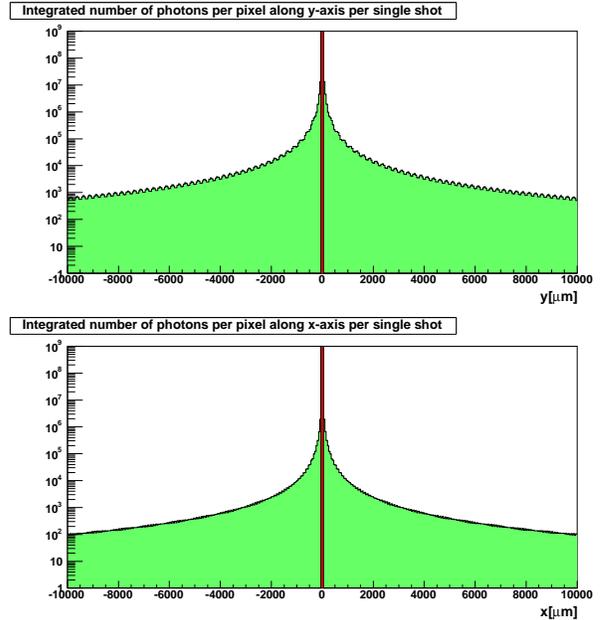}
\caption{
The integrated number of photons over each $50 \times 50 \mu$m${}^2$ pixel
area along $y$~(top) and $x$-axis~(bottom) at the focal plane per single shot
focusing. These correspond to the expected photon yields
in a single pixel column along the $y$ and $x$-axis in Fig.\ref{Fig4}.
The plotted region is further extended to $\pm 1$~cm in both axises.
All relevant laser parameters
are summarized in Tab.\ref{Tab1}. The ideal Gaussian pedestal distributions
without physical phase shift are also superimposed on the same figures.
The number of photons at the focal spot are truncated
to zoom the photon yield due to phase shift. By sampling photons
in regions sufficiently distant from the focal spot, one can increase
the signal-to-pedestal ratio.
}
\label{Fig5}
\end{figure}

The procedure to determine $\kappa$ by the physical interaction is as follows.
First, we determine constant phase biases $\phi_{opt}(X_i)$ caused by
all optical elements included in both SP and CP
by performing the phase contrast Fourier imaging at the ends of CP and SP.
We may assume that $\phi_{opt}(X_i)$ is stable over many pulse injections.
The phase $\phi_{opt}(X_i)$ can be measured with a CW laser
over a long period in between target-probe beam crossings during an experiment.
Since a less intense single mode Gaussian CW laser can provide a
homogeneous stable phase as the average value, we can, in principle, determine
$\phi_{opt}(X_i)$ with high accuracy based on the phase contrast Fourier imaging
integrated over the long time period at CP and SP by sharing the common
CW laser.
The template analysis discussed in \ref{subsecA} can also be applied to
this purpose. By assigning a square shape to the region $R_i$ representing
a pixel instead of the {\it signal} in (\ref{eq_chi2}),
we estimate $\kappa$ for each $X_i$.
The number of photons at a point $W_i$ on the focal plane contains
convoluted phase information on the amplitude from all points
on the transverse plane of the probe $N_X$ as seen from (\ref{eq_psiW}).
Therefore, as long as the number of sampling points on the focal plane $N_W$
is larger than that on the transverse probe profile $N_X$,
we can, in principle, determine a set of $\phi_{opt}(X_i)$
from (\ref{eq_chi2}) by scanning $\kappa$ for each $X_i$
over the expected range of the phase variation.
The achievable resolution of the phase reconstruction
depends on the scanning step on $\kappa$ in the $\chi^2$ test.
As we see later in Tab.\ref{Tab1}, Fig.\ref{Fig4}, and Fig.\ref{Fig5},
the phase contrast Fourier imaging at a focal plane achieves the
sensitivity to the order of $\sim 10^{-7}$ in the physical phase
shift by sampling only outer regions on the focal plane with
a several-photon-sensitive photo device. Therefore,
we can introduce the same resolution step to determine $\kappa$ for
$\phi_{opt}(X_i)$ as the analysis of the physical phase shift $\delta(X_i)$.
We may measure the initial coarse phase map $\phi_{opt}(X_i)$
{\it a priori} by a typical phase resolution $\sim \lambda/100$
with a commercially available wavefront sensor. In such a case
we need to repeat the two dimensional Fourier transform about $10^5$
times for scanning $\kappa$ for each $X_i$ to reach the same phase
resolution as $\sim 10^{-7}$ starting from the initial coarse resolution.
Accordingly, a proper computing power is necessary to solve the local
phases over a set of $\phi_{opt}(X_i)$.
Second, we can also apply the same procedure to determine a
set of $\phi_{pls}(X_i)$ on the pulse-by-pulse basis after mapping the
static phase $\phi_{opt}(X_i)$ without the physical template
$\delta(X_i)$ within CP.
Finally, we create a phase map by adding $\phi_{opt}(X_i)$ and
$\phi_{pls}(X_i)$ within SP. We then determine $\kappa$ caused
by the physical template $\delta(X_i)$ based on (\ref{eq_chi2})
with the same phase resolution step
when we determined $\phi_{opt}(X_i)$ and $\phi_{pls}(X_i)$.
\begin{table*}[t]
\begin{tabular}{lr}
\hline\hline
{\bf Target laser parameters} & {\bf Probe laser parameters} \\ \hline\hline
$\tau_t = 10$~fs & $\tau_p = {z_R}_t/c = 12$~fs\\ \hline
$E_t=10\mbox{kJ}$ ( $4.03\times10^{22}$ photons )&
$E_p=10\mbox{kJ}$ ( $4.03\times10^{22}$ photons )\\ \hline
$\lambda_t = 800$~nm & $\lambda_p = 800$~nm \\ \hline
$F_t = 1.2$ & $F_p = 4.5$  \\ \hline
${w_0}_t=F_t\lambda_t=0.96\mu$m & ${w_0}_p=F_p\lambda_p=3.6\mu$m\\ \hline
${z_R}_t=\pi{w^2_0}_t/\lambda_t = 3.6\mu$m &
${z_R}_p=\pi{w^2_0}_p/\lambda_p = 50.9\mu$m \\ \hline\hline
{\bf Embedded physical phase based on QED}\\ \hline\hline
$\delta = 3.17 \times 10^{-7}$ from (\ref{eq_delta_exp}) \\
($\delta n_{qed} = 1.34 \times 10^{-8}$ from (\ref{eq_delta_para}) 
with $\zeta =7$ and $\theta = \pi/2$) \\ \hline\hline
{\bf Focusing parameter in Signal Path} \\ \hline\hline
expansion factor $R_e = 5.0 \times 10^4$ \\ \hline
${{w}_p}_{e} = R_e {w_0}_p = 18$cm \\ \hline
${f_p}_{e} = 5\mbox{m}$ \\ \hline
Peak intensity ${{A^2}_p}_{e} = E_p / (2\pi {{w^2}_p}_{e})$
( $1.96 \times 10^{11}$ photons ) \\ \hline
\end{tabular}
\caption{
Parameters to produce Fig.\ref{Fig4} and \ref{Fig5} based on the conceptual
experimental setup in Fig.\ref{Fig3}.
The subscripts $t$, $p$ and $p_{e}$ denotes parameters associated with the
target laser, the probe laser and the probe laser after the expansion
before the final focusing in the Signal Path in Fig.\ref{Fig3}, respectively.
The definitions of parameters are explained in the text in
section~\ref{sec3} and subsection~\ref{subsecB}. As a basic constraint
on the common beam diameter $d$ of target and probe lasers in Fig.\ref{Fig3},
we assumed $d \sim 1$~m. 
}
\label{Tab1}
\end{table*}

As a demonstration,
Fig.\ref{Fig4}(top) illustrates the intensity distribution
on the focal plane in SP with an arbitrary unit on the contour height
in log-scale by sampling values with $50\mu$m steps in both
$x$ and $y$-axes. In this figure, only the physical
template $\delta(X_i)$ is embedded.
The parameters used for Fig.\ref{Fig4} are summarized
in Tab.\ref{Tab1}, where the parameters for both target and probe lasers,
the embedded physical phase shift due to the nonlinear QED effect,
and focusing parameter after the beam expansion to obtain the Fourier
transformed intensity distribution in SP are specified.
Figure \ref{Fig4}~(bottom) shows the same distribution without the physical
phase shift in the 3-D plot, where the pedestal Gaussian distribution is well
confined within an $20 \times 20\mu$m${}^2$ area.

Figure \ref{Fig5}~(top) and (bottom) show the integrated number of photons
in a single target-probe crossing per $50\mu$m $\times$ $50\mu$m pixel along
the $y$ and $x$-axes on the focal plane, respectively.
These correspond to the photon yield per camera pixel in single pixel line
along the $y$ and $x$-axes, respectively in Fig.\ref{Fig4}.
The plotted range is extended over $\pm 1$~cm.
In each of Fig.\ref{Fig5}, the number of photons in the peak region
confined in a pixel without phase shift is overlaid.
This indicates that the Gaussian pedestal part is well confined and
the modulation from the physical phase shift is expanded to
the outer directions on the focal plane. Therefore,
by sampling only the peripheral surrounding the focal point on the focal
plane without touching the most intense pedestal, we should be able to
count the sufficient number of photons per pixel on the pulse-by-pulse basis.
We may introduce a several-photon sensitive photo device with an
electrical amplification process such as the two-dimensional image intensifier
camera. This device can increase the sensitivity to the number of
countable photons and results in a higher resolution on the phase measurement.
The peripheral sampling within 1cm is possible by locating flexible
fiber bundles much like endoscopes with the $\mu$m precision,
as already performed in a test experiment
for another purpose\cite{Nondestructive}.

\subsection{Background in the phase contrast Fourier imaging}\label{subsecC}
The dominant background source of the current measurement may be caused by
the refractive index shift due to the plasma creation
from the residual gas along the path of the focused target laser pulse.
The refractive index of the static plasma in the limit of negligible collisions
between charged particles is expressed as
%
\begin{eqnarray}\label{EqNplasma}
N = \sqrt{1-\frac{\omega_p{}^2}{\gamma \omega_0{}^2}},
\end{eqnarray}
%
where
      $\omega_0$ is the angular frequency of the target laser,
      $\omega_p$ is the plasma angular frequency defined as
      $\sqrt{4\pi e^2 n_e / m_e}$ and
      $\gamma$ is the relativistic Lorentz factor
      given as $\sqrt{1+a_0^2}$ with
      $a_0=0.85\times10^{-9}\lambda[\mu m]\sqrt{I_0[W/cm^2]}$.
In the low pressure limit of the residual gas the amount of refractive
index shift $\Delta N$ is expressed as $\omega_p{}^2 / 2\gamma \omega_0{}^2$.
Although the refractive index in the plasma becomes smaller than that of
the peripheral area with neutral atoms, the inverted phase
contrast of the phase shift inside the probe pulse still maintains the
rectangular shape along the trajectory of the target laser.
Therefore, it should produce the same characteristic diffraction
pattern on the focal plane to that of the nonlinear QED case
as expected from the Babinet's principle.
In order to reduce this effect, we need to reduce the electron density
$n_e$ in the residual gas. If we take $\gamma \sim 1$ as the upper limit
of $\Delta N$ estimate, the refractive index shift
$\sim 10^{-11}$ due to the nonlinear QED effect for a reference
energy density $\sim 1$J/$\mu$m${}^3$,corresponding to the air pressure $\sim 10^{-6}$~Pa.
This pressure is easily attainable with conventional
vacuum pumps. The collisional frequency due to interactions between electrons
and ions is expected to be $10^8-10^9$s${}^{-1}$ at the critical electron
density $n_{cr}[cm^{-3}]=1.12\times 10^{21}/\lambda^2[\mu m]$ where $\omega_p$
equals $\omega_0$. For duration time of $\sim$fs of the target laser pulse,
the inverse bremsstrahlung radiation due to the collisional process
in the residual gas is negligible at $\sim 10^{-6}$~Pa.
Therefore, the dominant background contribution from the residual gas plasma
can be suppressed with the pressure well below $10^{-6}$~Pa.

We note that what actually happens is rather more dynamical
due to the pondermotive force by the high intensity laser field.
In such a case the refractive index shift based on the static plasma state
gives only the upper bound on the amount of the local refractive shift.
As discussed in \ref{subsecB}, we aim at the measurement of the refractive
index shift on the order of $\sim 10^{-8}$
due to the nonlinear QED effect in the intensity assumed in Tab.\ref{Tab1}.
This number is much greater than $\sim 10^{-11}$, as discussed above. Therefore,
even if the dynamical effect is taken into account, it should be negligible
to the measurement of the local refractive index shift,
because the absolute value of the refractive index shift is small enough.

\section{Probing Vacuum Fields in Very Low Energies}\label{sec4}
Extending on the QED investigation in the previous section,
we now wish to explore the nonlinear vacuum response to lasers mediated
by 'vacuum fields' whose energies are much lower than the energy of the pump
(and probe) lasers.
To the current knowledge no such fields have ever been observed.
This is in spite of the fact that there are numerous theories advanced.
These theories include: axions\cite{PDG},
minicharged particles\cite{MCP0,MCP1,MCP2,MCP3},
and dark energy\cite{DEreview, FujiiScalarTensor}.
Why is this? This may be thought of as follows.
After Rutherford's discovery of the inner core (i.e. nucleus) of an atom
being very tiny compared with the size of the already tiny size of the atom,
the experimental search went to explore ever smaller constituents of matter and
thus the thrust for higher energy or momentum experiments.
Theories have gone hand-in-hand
with this exploration, succeeding in ever shorter ranged interaction
theories and unification of forces, typified by the electroweak theory
\cite{Weinberg}. We have referred to
this standard and extremely successful method
as the high momentum approach. Almost all laboratory
research efforts have been on this approach to date.
This approach, though successful in exploring high energy
physics, is not suitable to explore energies much lower than eV.
As mentioned in Sec.\ref{Intro}, these fields that might exist in much lower
than eV cannot strongly couple to matter,
because if it did, it would have been long
discovered in lower energies. Thus these fields, if they ever exist, must
couple weakly. This means that we need to have an extremely strong driver to
manifest a sufficient signal overcoming this weakly coupling
interaction to show up above noise.
We have not possessed sufficiently powerful such photon
sources to date. Perhaps this may be changing now, however, with more intense
laser tools are to become available~\cite{MourouRMP}.
What we have called the high amplitude or high field
method~\cite{TajimaHighFieldScience, TajimaEPJD}
may provide an alternative path to detect such low energy weakly
interacting fields that are spread over semi-macroscopic scales.

When the constituent's energy is much lower than the probing photon's, it is
our suggestion that we employ two laser beams in the co-propagating geometry.
As we have motivated our method in the introduction, two co-parallel beams
produce a very low center of mass energy interaction at the beat frequency
(equal to the subtraction of the two laser frequencies)~\cite{Beat}.
The approach is a standard Brillouin scattering of laser in matter~
\cite{Brill, Chiao, Maradudin, Nishikawa, Rosen, Kruer, LLERev, McKinstrie}.
In media such as condensed matter and plasma the utility of these two points
have been recognized well in the study of nonlinear optics.
For example, the Brillouin forward scattering (BFS) method~\cite{TajimaLPBeam}
relies on two co-propagating laser beams with frequencies
very close to each other, where the difference between 
these two is matched with the eigenfrequency of the
acoustic modes of the medium, typically one of the lowest eigenmodes.
(We note that BFS is out of a more broad class of physical processes
of the parametric instabilities~\cite{Nishikawa}. In addition
a class of interaction under the modulational instabilities~\cite{MI}
shares a similar feature we may need.)
It was known that this process allows strong coupling of photons to low
frequency eigenmodes of the medium.
This interaction could resonate with the constituent's very low eigenfrequency,
should there be an eigenmode in its vicinity.
Second of all, the co-propagating setup
allows us to make the two beams interact over much
prolonged interaction time, thus much amplifying the nonlinearities and signal
arising from these.

\begin{figure}
\includegraphics[width=1.0\linewidth]{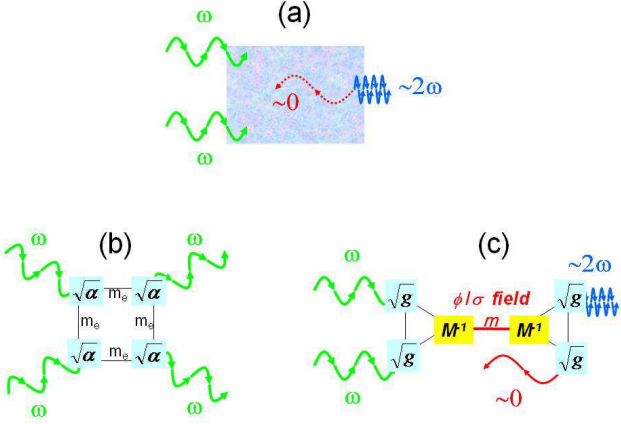}
\caption{
Schematic diagrams of photon-photon interactions in matter and in vacuum.
(a) Second harmonic generation in the experiment by Franken et al.
~\cite{Franken}, arising from the nonlinearity of crystal fields irradiated
by (intense enough) laser fields;
(b) Probing the QED vacuum nonlinearities, as suggested by
Heisenberg and Euler~\cite{EH,Weiscop},
where the vertices of the coupling in the Feynman diagram are characterized
by the fine structure constant of vacuum $\alpha$
and thus a weak nonlinearity requiring us to employ much more intense
fields than the case (a). The leading order interaction is the elastic
photon-photon scattering, though as a higher order there exists a second
harmonic generation as well; (c) Probing potential light-mass $m$ fields
in vacuum with intense laser fields. The vertices are characterized by very
feeble couplings of $M^{-1}$ and $g$~\cite{FujiiScalarTensor}.
The expected second harmonic generation may be said
to be not dissimilar to case (a). In order to increase
the observable signal, suggestions have been made.
}
\label{Fig6}
\end{figure}

In order to pick up the experimental signal of strong coupling to
the long-range mode, however, we suggest using the second harmonic
generation unlike Brillouin scattering where the detection
is normally performed via the first order phonon-photon coupling.
The pioneering research by Franken et al.\cite{Franken}
detected the nonlinearity in a quartz crystal.
Two photons in co-propagation strongly interacted through the quartz fields
over the coherence volume to produce the second harmonic generation.
This process may be schematically looked upon as the case in Fig.\ref{Fig6}(a).
There the quartz nonlinearities have mixed two forward propagating photons
$(\omega)$ to produce a photon with $2\omega$ (and possibly another photon
with frequency $\sim 0$).
In the previous sections we have considered nonlinear
QED process (Fig.\ref{Fig6}(b)). In this two incoming photons are mediated
by virtual electron-positron fields and outgoing are two photons.
The extreme forward scattering amplitude with quasi-parallel incident photons
is known to be largely suppressed in the QED process as we discuss later.
This is because the center-of-mass energy of the colliding two photons
is too low to satisfy the relevant mass scale of electron-positron pair.

In this section we consider a kinematically similar,
but much more daring measurement of the photon-photon interaction mediated
by low frequency "mode" of vacuum by setting up intense co-propagating lasers.
As we discuss later in detail, the process we focus is based on
Fig.\ref{Fig6}(c) which shows the second order in the photon-"mode" coupling.
It is as if we are extending forward Brillouin scattering with the strong
coupling to low frequency "mode" with the help of the resonance feature
and Franken's method~\cite{Franken} to obtain the clearer signature
to probe the second order photon-"mode" coupling.

%
%
In the previous sections, we have discussed the photon-photon interaction
based on the QED process projecting for a first experimental verification
of the real photon-photon interaction in the optical wavelength range.
In addition to the verification, the ratio between the first and the second
term in the one-loop effective Lagrangian in (\ref{eq_EHL})
can yield a general test to see whether vacuum contains other effects
beyond QED examining its value at 4:7.
Scalar field $\phi$ and pseudoscalar field $\sigma$ may contribute to
the first and second term, respectively.
Light scalar fields as a candidate of
dark energy have been recently intensively discussed~\cite{DEreview},
while the pseudoscalar fields (axion-like-particles) may be a source
of dark matter~\cite{PDG} and also possibly dark energy~\cite{deVega}.
However, we recognize that the test in the phase
contrast Fourier imaging is very limited in the mass and coupling
of those fields. In this section, therefore, 
we extend our method to search for those new
types of fields by instituting co-propagating laser beams.
These endeavors may be looked upon as in Fig.\ref{Fig6}(c).
Again two parallel photons come in, while two parallel photons come out.
We note that our approach is similar to but distinct from
many laboratory experiments with lasers
~\cite{BFRT, PVLAS, BMV, ALPS, LIPSS, OSQAR, GammeV}
already performed and proposed to search for those light fields.

%
%
Consider more details of the effective interaction Lagrangian as illustrated
in the triangle part in Fig.\ref{Fig6}(c), where $\phi$ and $\sigma$ couple
to the electromagnetic field via quantum anomaly-type
couplings~\cite{FujiiScalarTensor}
\beqa\label{eq_phisigma}
-L_{\phi}=g_{\phi} M_{\phi}^{-1} \frac{1}{4}F_{\mu\nu}F^{\mu\nu} \phi,
\nnb\\
-L_{\sigma}=g_{\sigma} M_{\sigma}^{-1}
\frac{1}{4}F_{\mu\nu}\tilde{F}^{\mu\nu} \sigma.
\label{mxelm_1}
\eeqa
Here $gM^{-1}$ denoted by subscript $\phi$ and $\sigma$
for scalar and pseudoscalar fields, respectively
provides the coupling strength. The dimensionless coupling
$g$ is typically proportional to the dimensionless
fine structure constant $\alpha$ for the two photons to couple to the
virtual charged particle pair in the triangle part. The effective coupling
includes the large mass scale $M$ to couple to the light field
with the mass $m$.
The large $M$ induces the weakness of the coupling via $M^{-1}$.
For example, the Newtonian constant $G$ is expressed as
$8\pi G = \hbar c M^{-2}_P$, where $M_P$ is the Planckian mass of $10^{27}$~eV.
The weakness of $G$ is the manifestation of the large mass scale at
the vertex in the triangle coupling.
In what follows we abbreviate the subscripts $\phi$ and $\sigma$ on
the coupling $gM^{-1}$ and the mass of light field $m$
unless we need to explicitly distinguish the type of the fields.
We use natural unit $\hbar = c = 1$
throughout subsequent sections, unless we explicitly note.

%
%
As a quite challenging case
we have attempted a theoretical approach to search for an
extremely light scalar field via the resonance in \cite{DEsearch},
which may be sensitive to the mass scale of $m_{\phi} \sim 10^{-9}$~eV
with $M^{-1}_{\phi} = M_P^{-1}$ corresponding to the gravitational coupling.
The method provides a new window into scoping the physics in the
Planckian mass scale by photon interactions in the quasi-parallel
incident laser beams in laboratories.
In this section we review the essence of the approach and further
develop basic formulae to apply the method to a concrete experimental
setup in order to discuss reachable limits on the coupling
strength $gM^{-1}$ and the mass $m$ for a given laser intensity
in the next section.

\begin{figure}
\includegraphics[width=1.0\linewidth]{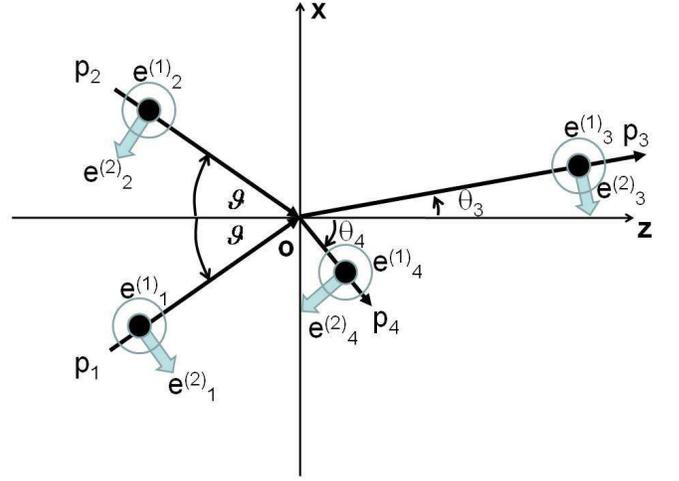}
\caption{
Definitions of kinematical variables for the suggested co-propagating
photons (this figure is quoted from \cite{DEsearch}).
}
\label{Fig7}
\end{figure}

%
%
As illustrated in Fig.\ref{Fig7},
we introduce an unconventional coordinate
frame in which two photons labeled by 1 and 2 sharing the same frequency are
incident nearly parallel to each other, making an angle $\vartheta$ with
the common central line along the $z$ axis. We define the $z-x$ plane
formed by $\vec{p}_1$ and $\vec{p}_2$.
The components of the 4-momenta of the photons are given by
$p_1 =(\omega\sin\vartheta,0,\omega\cos\vartheta ; \omega)$ and
the same for $p_2$ but with the sign of $\vartheta$ reversed, and
$p_3 =(\omega_3 \sin\theta_3, 0, \omega_3 \cos\theta_3 ; \omega_3)$ and
$p_4$  with $\omega_3, \theta_3$ replaced by $\omega_4, -\theta_4$,
respectively.
The angles $\theta_3$ and $\theta_4$ are defined also as shown
in Fig.\ref{Fig7}. This coordinate system can be transformed to
the center-of-mass (CM) system for the head-on collision ($\vartheta =\pi/2$)
by the Lorentz transformation
with $v/c\rightarrow 1$ for $\vartheta \rightarrow 0$.
Conversely, this implies that the realization of the quasi-parallel collision
in the laboratory frame corresponds to the realization of the
extremely low CM energy, as we see below.

In this frame one of the final photons in the forward direction along
the $z$ axis must have an upshifted frequency due to the energy-momentum
conservation independent of the physical origin of the dynamics.
In the limit of $\vartheta \rightarrow 0$, a process of
$\omega_3 \rightarrow 2 \omega$
is realized. This frequency doubling nature is an extremely valuable
characteristics from the experimental point of view,
as compared to the case with no frequency shift in the
center-of-mass system. In addition, more importantly,
it is essential to maintain a quasi-parallel nature of the incident
beams in order to access the resonance point, as we shall stress later.

The energy-momentum conservation laws are
\beqa
0 \mbox{-axis}:&&\omega_3 +\omega_4 = 2\omega, \label{kinm_3}\\
z\mbox{-axis}:&&\omega_3 \cos\theta_3 + \omega_4 \cos\theta_4=
2\omega \cos\vartheta, \label{kinm_4}\\
x\mbox{-axis}:&&\omega_3\sin\theta_3 =\omega_4\sin\theta_4.
\label{kinm_5}
\eeqa
From the conditions $0<\omega_{3,4} <2\omega$, we may choose
$0<\theta_3<\vartheta<\theta_4<\pi$,
without loss of generality.  From \reflef{kinm_3})-\reflef{kinm_5})
we derive the relation
\beq
\sin\theta_3 =\sin\theta_4\frac{\sin^2\vartheta}
{1-2\cos\vartheta\cos\theta_4+\cos^2\vartheta}.
\label{kinm_7}
\eeq

The differential elastic scattering cross section favoring the
higher photon energy $\omega_3$ is given by
\beq
\frac{d\sigma}{d\Omega_3}=(8\pi \omega)^{-2}
\sin^{-4}{\vartheta} (\omega_3/2\omega)^2 |{\cal M}|^2,
\label{kinm_13}
\eeq
where ${\cal M}$ is the invariant amplitude and
\beq
\omega_3 =\frac{\omega \sin^2\vartheta}{1-\cos\vartheta \cos\theta_3}.
\label{kinm_8}
\eeq
Here we expect the upshifted frequency $\omega_3 \rightarrow
2\omega$, as $\theta_3\rightarrow 0$ for  $\vartheta \rightarrow 0$, as
mentioned before.

%
%
%
%
The resonance decay rate of light fields with the
mass $m$ into two photons is expressed as
\beq
\Gamma=(16\pi)^{-1} \left( g M^{-1}\right)^2 m^3.
\label{mxelm_4a}
\eeq
The light field is exchanged between the pairs
$(p_1, p_2)$ and $(p_3, p_4)$, thus giving the squared four-momentum
of the scalar field
\beq
q_s^2 =\left(p_1+p_2\right)^2 =2\omega^2 \left( \cos 2\vartheta -1
\right)
\label{mxelm_4b}
\eeq
with the metric convention $(+++-)$ for the definition of four momenta.

With the polarization vectors given by
$\vec{e}_i^{\ (\beta)}$ where $i=1, \cdots,4$ are the photon labels, whereas
$\beta =1,2$ are for the kind of linear polarization as depicted in
Fig.\ref{Fig7}, we summarize the non-zero invariant amplitudes for scalar
exchanges
\beq\label{eq_M11}
{\cal M}_{1111}={\cal M}_{2222}=-{\cal M}_{1122}=- {\cal M}_{2211},
\eeq
and for pseudoscalar exchanges
\beq\label{eq_M12}
{\cal M}_{1212}={\cal M}_{1221}=-{\cal M}_{2112}=- {\cal M}_{2121},
\eeq
where the first two digits in the subscripts
correspond to the states of the linear polarization
of incoming two photons 1 and 2, respectively and the last two correspond
to those of outgoing two photons 3 and 4, respectively
as illustrated in Fig.\ref{Fig7}.

We focus on one of these non-zero amplitudes by denoting it as ${\cal M}$;
\beq
{\cal M} =-(g M^{-1})^2\frac{\omega^4 \left(
\cos2\vartheta -1\right)^2}{2\omega^2 \left(
\cos2\vartheta -1\right)+m^2},
\label{mxelm_7}
\eeq
where the denominator, denoted by ${\cal D}$, is the light field propagator.
We note $q_s$ is timelike.  We then make the replacement
\beq
m^2 \rightarrow \left( m -i\Gamma \right)^2 \approx
m^2 -2im \Gamma.
\label{mxelm_9}
\eeq
Substituting this into the denominator in \reflef{mxelm_7}), and
expanding around $m$, we obtain
\beq
\hspace{-.1em}{\cal D}\approx -2\left( 1-\cos2\vartheta  \right) \left( \chi+ia
\right),\quad\hspace{-.7em}\mbox{with}\quad\hspace{-.7em} \chi =\omega^2 -\omega_r^2,
\label{mxelm_10}
\eeq
where
\beq
\omega_r^2 =\frac{m^2/2}{1-\cos 2\vartheta },\quad
a=\frac{m \Gamma}{1-\cos 2\vartheta}.
\label{mxelm_12}
\eeq
From \reflef{mxelm_4a}) and \reflef{mxelm_12}),  $a$ is also expressed as
\beq\label{eq_a}
a = \frac{\omega^2_r}{8\pi}\left(\frac{g m}{M}\right)^2,
\eeq
which explicitly shows the proportionality to $M^{-2}$.
We then finally obtain the expression for the squared amplitude as
\beq
|{\cal M}|^2 \approx  (2\pi)^2 \frac{a^2}{\chi^2+a^2}.
\label{mxelm_13}
\eeq
As for the off-resonance case $\chi\gg a$, $|{\cal M}|^2$
is largely suppressed due to the factor $a^2 \propto M^{-4}$ for
the case of small coupling $M^{-1}$.
On the other hand, if experiments take ideally the limit of
$\omega \rightarrow \omega_r$,
$|{\cal M}|^2 \rightarrow (2\pi)^2$ is realized from (\ref{mxelm_13}).
This is independent of the smallness of the factor $M^{-4}$
as expected from the off-resonance case or equivalently
from the square of (\ref{mxelm_7}).
This is the most important feature arising from the resonance that overcomes
the weak coupling stemming from the large relevant mass scale such as $M=M_P$.
However, we then confront the extremely narrow width $a$
for {\it e.g.} $gm \ll 1$~eV , $M \sim M_P=10^{27}$~eV and
$\omega_r \sim \mbox{1 eV}$.
We now discuss how to overcome this difficulty.

In conventional high energy collisions the beam momenta are implicitly
supposed to be their mean values. This is because the momentum spread,
or the uncertainty expected from the de Broglie wavelength of the
relativistic particle is negligibly small compared to the
relevant momentum exchanges in the interaction
that experiments are interested in.
Resonance searches in such experiments must adjust $\chi$ in (\ref{mxelm_10})
and (\ref{mxelm_13}) so that the mean $\chi$ is close enough to the
peak location within $\pm a$.
On the other hand, in the case of the co-propagating laser
beam aiming at detection of extremely small momentum exchanges via
the resonance, the situation is quite different
due to the nature of incident waves. This is because
the uncertainty included in the initial photon momenta
is much larger than the relevant energy scale of the resonance,
leading to the condition $|\chi| \gg a $.
In this case the squared scattering amplitude must be integrated
over the possible uncertainties on the incident
wave function on the event-by-event basis.
As we will explain below, the uncertainty in the incident photon momenta
is related to the uncertainty in $\chi$ via the uncertainty in the incident
angle $\vartheta$. Therefore, it is instructive to consider the feature
of the integral of the resonance function
of the Breit-Wigner(BW) formula~\cite{BW} from $\chi_-$ to $\chi_+$ as follows;
\beq\label{eq_IntBW}
I = \int_{\chi_-}^{\chi_+}  \frac{a^2}{\chi^2+a^2} d\chi
  = \left[a\tan\left(\frac{\chi}{a}\right)\right]_{\chi_-}^{\chi_+},
\label{mxelm_14}
\eeq
where $I=a\pi/2$ and $I=a\pi$ for $\chi_+=-\chi_-=a$ and
$\chi_+=-\chi_-=\infty$, respectively.
This indicates that the value of integral is proportional to $a$, {\it i.e.}
$M^{-2}$ from (\ref{eq_a}).
The value ranges for the finite and infinite integrals over
only factor of two. From this fact, we expect that
the integral enhances the squared scattering amplitude by a factor of $M^2$
compared to the non-resonance interaction proportional to $M^{-4}$ from
(\ref{mxelm_7}), as long as the peak is contained within the experimental
resolution on $\chi$, {\it i.e.} the condition $\chi_+ > a$ and $\chi_- < -a$
is satisfied.
This implies that experiments in the case of the co-propagating laser beam
configuration do not have to make efforts to adjust $\chi$ close to the
extremely narrow resonance region thanks to the consequent huge enhancement
by the integral over the wide range on $\chi$.
Meanwhile, it is difficult to identify
the exact location of the resonance mass within the wide gate on $\chi$.
Since we are interested in having the sensitivity to the extremely weak
coupling such as gravity $M=M_P$, the enhancement of the squared amplitude
is more crucial than finding the exact location of resonance masses.
As we discuss in the following sections, however,
we may be able to provide a crude estimate on the order of
the mass scale of the resonance even in such a situation.

The consideration above leads us to parametrize the squared scattering amplitude
as follows:
\beqa\label{eq_M2ave}
\overline{|{\cal M}|^2}=\int_{-\infty}^{+\infty} \rho(p_1, p_2)
|{\cal M}|^2 d\chi
\quad \qquad \qquad \qquad \nnb\\
= (2\pi)^2\int_{-\infty}^{+\infty}\rho(p_1, p_2)
\frac{a^2}{\chi^2(p_1,p_2)+a^2}d\chi,
\label{mxelm_15}
\eeqa
where $\rho(p_1, p_2)$ is the normalized probability distribution
to supply nominal combinations $(p_1, p_2)$ from incident laser fields
and the real part $\chi(p_1,p_2)$ is indirectly specified by $(p_1, p_2)$
via the incident angle $\vartheta$ between incident two photons.
This parametrization expresses averaging over
possible combinations of $p_1$ and $p_2$.
We note that $(p_1, p_2)$ is not {\it a priori} observed momenta,
but just a nominal specification among the possible initial momenta.
In other words, $\rho(p_1, p_2)$ is not a statistical weight on the
discrete momenta after the contraction of the wavepacket of each photon state.
This implies that an infinite statistics is not necessary to obtain
the continuous nature. Rather, we need this treatment even for
a two photon state as long as the source of photons is
not the perfect plane wave.
This allows the continuous integral on $\chi$ via the continuous combination of
$\rho(p_1, p_2)$ in (\ref{eq_M2ave}).
As long as the probability weight $\rho(p_1, p_2)$ is close enough to
unity around the resonance peak, the enhancement discussed
with (\ref{eq_IntBW}) is guaranteed. This is the essence of our main
strategy of the co-propagating configuration to overcome
the difficulty due to the narrow resonance width $a$.

In order to design experiments, we start from the resonance condition,
the first of \reflef{mxelm_12}), by assuming $\vartheta \ll 1$,
\beq\label{eq_CMreso}
m \sim 2\vartheta \omega.
\eeq
We note that the product $2\vartheta\omega$ corresponds to the CM energy
of incident two photons.
This indicates that experiments have the two adjustable handles for a given
mass scale or the CM energy.
We emphasize that we can lower the CM energy by several orders
of magnitude by only introducing smaller $\vartheta$ with fixed $\omega$.
This should be contrasted to high energy colliders where we need a large
amount of efforts to increase the CM energy by an order of magnitude.
This advantage is also supported from a technical point of view,
since scanning the incident angle $\vartheta$ should be much easier than
scanning the energy $\omega$ of the resonance.
We point out that the resonance condition (\ref{eq_CMreso}) is not just at
one point, but rather in a hyperbolic band in the $\vartheta-\omega$ plane
given with a finite resolution with $\delta\vartheta$ of
the incident angle $\vartheta$.
This implies that the deviation $\delta\omega$ from the resonance energy
$\omega_r$ can satisfy the same resonance condition with a different $\vartheta$
within $\pm \delta\vartheta$.
As far as $\delta\omega/\omega \ll \delta\vartheta/\vartheta$ is satisfied
in the setup, we can ignore the effect of $\delta\omega$.
This is in fact the case, as seen in the following discussions.
Therefore, we can take the attitude that we fix the incident energy
at the optical frequency $\omega = \omega_{opt}$
and scan $m$ by changing $\vartheta$ around $\vartheta_r$,
where $\omega_{opt}$ and $\vartheta_r$ satisfy the resonance condition
based on (\ref{eq_CMreso})
\beq\label{eq_omegaopt}
\omega^2_{opt} = m^2/(4\vartheta^2_r).
\eeq
In this case the variation on $\vartheta$ leads to
the variation on $\chi$. This is expressed by the following relation
based on the second of (\ref{mxelm_10})
\beq\label{eq_the2x}
\chi(\vartheta) = \omega^2_{opt} - \frac{m^2}{4\vartheta^2}
= \omega^2_{opt} (1-\varepsilon^{-2}),
\eeq
where
$\varepsilon \equiv \vartheta / \vartheta_r$ in the unit of $\vartheta_r$
is introduced.

We now discuss the average of the squared amplitude $\overline{|{\cal M}|^2}$
over the possible uncertainty on the incident angle $\vartheta$
\beqa\label{eq_the2eps}
\overline{|{\cal M}|^2} &=& 
\int^{\pi/2}_{0} \rho(\vartheta) |{\cal M}|^2 d\vartheta
\quad \qquad \nnb\\
&=& \int^{\pi/(2\vartheta_r)}_{0} \rho(\varepsilon) 
|{\cal M}|^2 \vartheta_r d\varepsilon,
\eeqa
where $\rho(\vartheta)$ is a probability distribution function
normalized between 0 and $\pi/2$
as a function of continuous uncertainty on $\vartheta$
from arbitrarily chosen two photon combinations within a laser pulse.
The incident angle $\vartheta$ is re-expressed with $\varepsilon$
with $d\vartheta = \vartheta_r d\varepsilon$ for the second of (\ref{eq_the2eps}).
From the relation (\ref{eq_the2x}), we find $\varepsilon = (1-x)^{-1/2}$ and
$d\varepsilon = 1/(2\omega^2_{opt}) \varepsilon^3 dx$.
Equation (\ref{eq_the2eps}) is then re-expressed with $\chi$
\beqa\label{eq_eps2x}
\overline{|{\cal M}|^2} = (2\pi)^2 \frac{\vartheta_r}{2\omega^2_{opt}} 
\qquad \qquad \qquad \qquad \qquad \qquad \nnb\\
\int^{1-(2\vartheta_r/\pi)^2}_{-\infty}
\frac{\rho((1-\chi)^{-1/2})}{(1-\chi)^{3/2}} \frac{a^2}{(\chi^2+a^2)} d\chi,
\quad
\eeqa
where (\ref{mxelm_13}) is substituted.
This equation is the exact representation of (\ref{eq_M2ave}) starting from
the uncertainty on the incident angle $\vartheta$, if the upper limit
of the integral range is regarded as large enough compared to $a$.
Let us define $x \equiv a\xi$ to explicitly discuss the structure of the
integral kernel in the unit of the width $a$ of BW.
With $\xi$, (\ref{eq_eps2x}) is further re-expressed as
\beqa\label{eq_x2xi}
\overline{|{\cal M}|^2} =
(2\pi)^2 \frac{\vartheta_r}{2\omega^2_{opt}} a
\qquad \qquad \qquad \qquad \qquad \qquad \nnb\\
\int^{a^{-1} \{1-(2\vartheta_r/\pi)^2 \}}_{-\infty}
\frac{\rho((1-a\xi)^{-1/2})}{(1-a\xi)^{3/2}} \frac{1}{\xi^2+1} d\xi,
\eeqa
where the first factor of the integral kernel corresponds to a normalized
weight function and the second is BW with the width of unity.
This expression explicitly shows the enhancement by the factor of $a$ implying
the proportionality to $M^{-2}$ based on (\ref{eq_a}).
As long as $\rho$ is a monotonic function, the weight function in front of
BW can be close to unity for small $\xi$ because of $a\xi \ll 1$.
With such a weight we expect that the value of the integral may be
close to that of the BW as we discuss with a concrete weight function
in the following section.

Remaining issue is how to further cope with the problem of still very
small $M^{-2}$, though much larger than $M^{-4}$, in experiments.
First, this can be solved by $\sin^{-4}\vartheta$ behavior of the cross
section in (\ref{kinm_13}) that arises from the phase volume factor and
the flux factor in the quasi-parallel two photon interaction. For extremely
light mass, this factor gains a large number due to small $\vartheta_r$.
Second, the intense laser fields can provide a large luminosity and
the intensity of the signal is proportional to square of the intensity
of laser in the limited case of the incoherent two photon interaction.
We have three ingredients or knobs; the $M^2$ enhancement by the weighted
BW integral, the factor of $\vartheta_r^{-4}$ and
the growth of laser intensity. By marshalling these knobs,
we expect to increase the detectability for the undiscovered light fields
in vacuum, which have evaded from our grasp to date.

In the following sections we consider the experimental realizations
with $\omega_r \sim 1$~eV (optical laser) and $O(1)< \vartheta_r < O(10^{-10})$
by scanning of the mass range $O(1)< m < O(10^{-10})$~eV.
We then plug explicit weight functions into (\ref{eq_x2xi})
based on the suggested experimental setup. By combining (\ref{eq_x2xi}) and
(\ref{kinm_13}), we discuss reachable mass-coupling limits for a
given laser intensity attainable in future experiments.

\section{Conceptual laser measurement}\label{sec5}
We emphasized the importance of photon-photon interaction in a co-parallel
or at small angle setup in order to increase the signal of detecting low
energy constituents.
Thus a simple way is to explore the mass range
$m<\pi\omega$ by using two independent laser beams with
the small incident angle.
We then directly measure the resonance curve in (\ref{eq_CMreso}) by
scanning both $\vartheta$ and $\omega$ to quantitatively
observe the nature of the resonance curve. For the very smaller mass scale,
or equivalently smaller incident angle,
however we must take into account the beam spread in the diffraction limit.
This determines the controllable smallest incident
angle, or the mass scales of the light fields we look for.
We consider here one-beam focusing in order to provide the simplest basis
to quantify reachable mass-coupling limits for given a set of experimental
parameters.

\begin{figure}
\bcent
\includegraphics[width=1.0\linewidth]{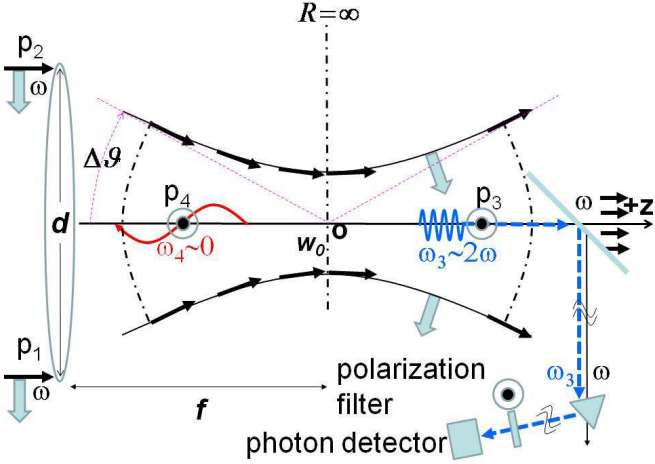}
\caption{
Suggested experimental setup of the co-propagating photon
interaction and detection. The linear polarizations of incident
and outgoing photons are drawn only for the scalar exchange
with the scattering amplitude $|{\cal M}_{1122}|$ as an example.
}
\label{Fig8}
\ecent
\end{figure}

The conceptual experimental setup with one-beam focusing geometry
is illustrated in Fig.\ref{Fig8}.
Incident photons from a Gaussian laser pulse with linearly polarization
are focused by the conceptual lens component into the diffraction limit.
Quasi-parallel incident photons interact with each other
between the lens and the focal point, from which photons 3 and 4 are emitted
nearly in the opposite direction along the $z$ axis
with $\omega_3\sim 2\omega$ and $\omega_4 \sim 0$.
The mirror with dichroic nature is transparent to the
non-interacting photons with the beam energy of $\omega$,
while $\omega_3$ is reflected to the prism
(equivalent to a group of dichroic mirrors)
which selects $\omega_3$ among residual $\omega$
and sends it to the photon detector placed off the $z$-axis.
This process is assisted by the polarization filter.
From the polarization dependence of the invariant amplitude in
(\ref{eq_M11}) and (\ref{eq_M12}), the combinations of polarizations of
two photons between the initial and final states must satisfy
$11 \rightarrow 11(22)$ for the scalar field exchange and
$12 \rightarrow 12(21)$ for the pseudoscalar field exchange.
We note that we can choose the type of fields we search for
by setting the initial polarization state. In the case of one-beam focusing
the search for the scalar field is easier, because we do not have to
mix the two polarization states as in the case of the pseudoscalar field.
Furthermore, the selection of the rotated final state $22$ can enhance the
signal-to-background ratio for the scalar field case, because a huge
number of non-interacting photons has the final polarization state of $11$.
In what follows we provide formulae to evaluate the accessible limit on the
mass-coupling defined in (\ref{mxelm_1}) for a given laser intensity,
in the case that we detect a double frequency photon per laser focusing.

%
%
The Gaussian laser parameters are summarized in (\ref{eq_Gauss})
through (\ref{eq_z0}). We now estimate the effective luminosity
${\cal L}$ over the propagation volume of the laser pulse.
We now restore physical dimensions of $\hbar$ and $c$ in this section,
unless we explicitly note.
Consider a Gaussian laser pulse with duration time $\tau$
with the speed of light $c$ and the
average number of photons $\bar{N}$ per pulse.
The exchange of light field may take place anywhere within the volume
defined by the transverse area of the Gaussian laser times the focal length $f$
before reaching the focal point.
We first consider the effective number of photons $N_{int}$ during
an interaction with the time scale of $\Delta t$.
As a result of the interaction we observe a frequency doubled photon
in the laboratory frame. The momentum transfer of $\sim \hbar\omega/c$
between photons defines the minimum interaction time scale from the
uncertainty principle as follows
\beqa\label{eq_Tint}
\Delta t > 2\pi \omega^{-1}.
\eeqa
The effective number of photons during $\Delta t$ is expressed as
\beqa\label{eq_Nint}
N_{int} = \frac{\Delta t}{\tau}\bar{N}.
\eeqa
Making the pulse duration $\tau \sim \Delta t$
maximizes the instantaneous luminosity.
Suppose a point $z$ along the laser propagation axis.
The instantaneous luminosity at the point $z$ is defined as
\beqa\label{eq_Linstant}
L(z) = \frac{C(N_{int},2)}{\pi w^2(z)}
\sim \frac{N^2_{int}}{2\pi w^2_0}\frac{z^2_R}{z^2+z^2_R}
\eeqa
where
$C(N_{int},2)$ denotes a combinatorics to choose two photons among
a large number of photons available within time scale of $\Delta t$ and
the expression of $w^2(z)$ in (\ref{eq_wz}) is substituted to
obtain the second with the approximation for the combinatorics.
We then consider the averaged instantaneous luminosity $\bar{L}$
over the focal length $f$ as follows;
\beqa\label{eq_Lbar}
\bar{L} = f^{-1} \int^f_0 L(z) dz \sim 
\qquad \qquad \qquad \qquad \qquad \nnb\\
\frac{N^2_{int}}{2\pi f w^2_0} z_R \tan^{-1}(f/z_R)
= \frac{N^2_{int}}{2 f \lambda} \tan^{-1}(f/z_R).
\eeqa
The number of effective bunches $b$ is related with $f$ as
\beqa\label{eq_b}
b = \frac{f}{c\Delta t}.
\eeqa
The effective luminosity ${\cal L}$ over the propagation volume of the
laser pulse is finally expressed as
\beqa\label{eq_L}
{\cal L} = b \bar{L}
\qquad \qquad \qquad \qquad \quad
\qquad \qquad \qquad \qquad \qquad \nnb\\
= \frac{f}{c\Delta t} \frac{N^2_{int}}{2f\lambda}\tan^{-1}(f/z_R)
= \frac{\Delta t}{\tau} \frac{\bar{N}^2}{2c\tau \lambda}\tan^{-1}(f/z_R),
\eeqa
where (\ref{eq_Nint}) is substituted in the last step.

The uncertainty on the incident angle between two light waves is expected to be
\beqa\label{eq_Delta}
\Delta\vartheta \sim
\frac{\lambda}{z_R} = \pi^{-1} \left( \frac{\lambda}{w_0} \right)^2
\eeqa
from the definition of the Rayleigh length $z_R = \pi w^2_0 / \lambda$.
The minimum beam waist $w_0$ at $z=0$ varies with the experimental conditions,
the focal length $f$, Rayleigh length $z_R$ and diameter of conceptual lens $d$,
\beq\label{eq_dfw0}
w_0 = (d/2)\frac{(f/z_R)}{\sqrt{1+(f/z_R)^2}}.
\eeq
We see that $\Delta\vartheta$ may be controlled via $w_0$ by choosing
suitable $f$ and $d$ in experiments.

The possible uncertainty on the incident angle $\vartheta$ affects
the average of the squared amplitude $\overline{|{\cal M}|^2}$ as shown in
the first of (\ref{eq_the2eps}).
In order to obtain an approximation close enough to reality,
we plug the following step function into (\ref{eq_the2eps}):
\beqa\label{eq_rho}
\rho(\vartheta) = \left\{
\begin{array}{ll}
1/\Delta\vartheta & \quad \mbox{for $0 < \vartheta \le \Delta\vartheta$} \\
0 & \quad \mbox{for $\Delta\vartheta < \vartheta \le \pi/2$}
\end{array}
\right\},
\eeqa
which is normalized to the physically possible range $0< \vartheta \le \pi/2$.
By substituting (\ref{eq_rho}) into (\ref{eq_x2xi}), we obtain
\beqa\label{eq_M2xi}
\overline{|{\cal M}|^2} =
\frac{(2\pi)^2}{2\omega_{opt}^2} \frac{\vartheta_r}{\Delta\vartheta} a
\qquad \qquad \qquad \qquad \qquad \quad \nnb\\
\int^{a^{-1} \{1-(\vartheta_r/\Delta\vartheta)^2 \}}_{-\infty}
\frac{1}{(1-a\xi)^{3/2}} \frac{1}{(\xi^2+1)} d\xi,
\eeqa
where the first factor of the integral kernel corresponds to
the weight function and the second is Breit-Wigner(BW) with the width of unity.
The weight function of the kernel is close to unity
for small $\xi$ due the smallness of $a$ in (\ref{eq_a}).
Therefore, the value of integral in (\ref{eq_M2xi}) is almost
equivalent to that of BW~\cite{footnoteBW}. This is because
the monotonic positive weight function approaches zero
as $\xi \rightarrow -\infty$ more rapidly than the pure BW,
whereas the pure BW suppresses the increase
of the weight function close to zero
at $\xi \rightarrow a^{-1} \{1-(\vartheta_r/\Delta\vartheta)^2\}$
for $\Delta\vartheta < 1$.
We then approximate (\ref{eq_M2xi}) as the integrated BW
over $\pm \sim \infty$ as follows:
\beqa\label{eq_M2final}
\overline{|{\cal M}|^2} \sim
\frac{(2\pi)^2}{2\omega_{opt}^2} \frac{\vartheta_r}{\Delta\vartheta} a \pi.
\eeqa
With $a$ in (\ref{eq_a}) and $\overline{|{\cal M}|^2}$ in (\ref{eq_M2final}),
the differential cross section in (\ref{kinm_13}) is finally expressed as
\beqa\label{eq_dSdO3}
\overline{\left(\frac{d\sigma}{d\Omega_3}\right)}
\sim
\frac{\pi}{64} \left( \frac{2\pi}{\lambda} \right)^{-2}
\left(\frac{\vartheta_r}{\Delta\vartheta}\right)
\left(\frac{g m}{M}\right)^2
\vartheta^{-4}_r,
\eeqa
where the approximations $\omega_3 \sim 2\hbar\omega$ and $\vartheta_r \ll 1$
are taken into account.

Multiplying (\ref{eq_L}) by (\ref{eq_dSdO3}),
we obtain the differential
yield per laser pulse focusing as follows:
\beqa\label{eq_dYdO3} \frac{d{\cal Y}}{d\Omega_3}
= {\cal L} \overline{\left(\frac{d\sigma}{d\Omega_3}\right)}
\quad \qquad \qquad \qquad \qquad \qquad \qquad \qquad \nnb\\
= \frac{1}{512\pi} \frac{\Delta t}{\tau} \frac{\lambda}{c\tau}
\tan^{-1}(f/z_R)
\left(\frac{\vartheta_r}{\Delta\vartheta}\right)
\left(\frac{g m}{M}\right)^2 \vartheta^{-4}_r \bar{N}^2.
\eeqa

There are several experimental knobs to affect the observable events in
(\ref{eq_dYdO3}).
If we choose $\tau \sim \Delta t \sim \lambda/c$ resulting in
$c\tau \sim \lambda$, we can maximize the effective luminosity.
From (\ref{eq_Delta}), the reduction of $\Delta\vartheta$ or increasing $w_0$
enhance the yield. From (\ref{eq_dfw0}), we express $f/z_R$ as
\beq\label{eq_f2z0}
(f/z_R)^2 = \frac{w^2_0}{(d/2)^2 - w^2_0}.
\eeq
From this relation, increasing $w_0$ as large as $(d/2)$ also enlarges
$f/z_R$. This introduces a slight increase for the factor $\tan^{-1}(f/z_R)$,
though its effect is tempered by the nature of $\tan^{-1}$.

As a short summary, we make the most important note from the experimental
point of view based on this conceptual design. The condition
$\vartheta_r/\Delta\vartheta = 1$ maximizes the chance
to search for the resonance, while $\vartheta_r/\Delta\vartheta > 1$ results
in a huge suppression of the cross section by $M^{-4}$ ($\hbar=c=1$)
as we discussed.
This is because the resonance peak is out of the region covered
by $\Delta\vartheta$. This parameter corresponds to the sharp cut-off
of the cross section. Therefore, controlling $\Delta\vartheta$ via relations
(\ref{eq_Delta}), (\ref{eq_dfw0}) and (\ref{eq_f2z0}) can provide an
experimental way to define the mass range that we eliminate
if no signal is found. From the resonance condition in (\ref{eq_CMreso}),
we evaluate the lower bound on the mass range we can exclude by this
conceptual design as follows ($\hbar=c=1$):
\beqa\label{eq_mbound}
m_{min} \sim 2\vartheta_r \omega_{opt} \sim 2\Delta\vartheta\omega_{opt}
\qquad \qquad \qquad \nnb\\
= 2\pi^{-1} (\lambda/w_0)^2\omega_{opt}> 2 \pi^{-1} (\lambda/d)^2\omega_{opt},
\eeqa
where (\ref{eq_dfw0}) is substituted and
the obvious experimental condition $w_0 < d$ is required for the
last inequality. Below this lower mass bound we cannot discuss what
the mass scales of the light fields are, even though the measurement is
still sensitive to the presence of lighter resonances than the lower mass bound
due to the large enhancement by $\vartheta^{-4}_r$.

\section{Discussion based on one-beam focusing}\label{sec6}
As a demonstration we now discuss the necessary laser intensity
for the following reference case based on the one-beam focusing setup.  First,
we note that $\Delta t$ in (\ref{eq_Tint}) is the resolvable minimum time scale.
As long as we discuss extremely low mass field, the interaction
time scale may be over $\hbar/mc^2$.
In the case of low mass the interaction time scale may be
much longer than $2\pi \omega^{-1}_{opt}$. On the other hand,
photon-photon interactions must occur within $\tau$.
This fact implies we should assume $\Delta t/ \tau = 1$
in the case $\Delta t > \tau$.
Since the effective luminosity tends to be enhanced for smaller beam waists,
we assume a short pulse laser with duration time close to
$2\pi \omega^{-1}_{opt}$ in the following discussion.
Let us assume that the diameter of the conceptual lens is $\sim 2$~m
by taking the damage threshold of the lens into account
for the use of high intense laser pulse.
With the assumption $f \ll z_R$, the minimum beam waist in
(\ref{eq_dfw0}) is approximated as $w_0 \sim (df\lambda/(2\pi))^{1/3}$.
This gives the cut-off of the sensitive mass region $m_{cut}$
defined by the condition $\vartheta_r < \Delta\vartheta$
via the relation (\ref{eq_Delta}) as follows;
\beq\label{eq_mrange}
m_{cut} \equiv 2\omega_{opt}\Delta\vartheta =
2\omega_{opt}\frac{\lambda^2}{\pi(df\lambda/(2\pi))^{2/3}}.
\eeq
If we are allowed to change the focal length between $\sim 1 < f < \sim 1000$~m,
the cut-off mass range varies from $\sim 10^{-9}$ to $\sim 10^{-11}$~eV
with $\omega_{opt}\sim 1$~eV.
By choosing the focal length realizable in laboratories as $f \sim 3$~m,
we expect $m_{cut} \sim 10^{-10}$~eV. For $\omega_{opt}\sim 1$~eV,
the resonant incident angle is $\vartheta_r \sim 10^{-10}$ from
(\ref{eq_CMreso}). As the most challenging case, we assume the coupling
as weak as gravity $M^{-1} \sim M^{-1}_P$.
We are now ready to estimate the average number of
photons $\bar{N_1}$ to expect
$d{\cal Y}/d\Omega_3 \sim 1$ per pulse focusing based on \reflef{eq_dYdO3})
for the physical parameters: $m \sim 10^{-10}$~eV,
$g \sim \alpha=1/137$ and $M_P \sim 10^{27}$~eV,
and for the experimental parameters discussed above:
$\omega_{opt} \sim 1$~eV,
$\tau \sim 10$~fs,
$\Delta t/ \tau = 1$,
$w_0 \sim 0.01$~m,
$d \sim 2$~m,
$f\sim 3$~m,
$\Delta\vartheta \sim 4 \times 10^{-9} > \vartheta_r$
and
$f/z_R \sim w_0$ from (\ref{eq_dfw0}) resulting
in $\tan^{-1}(f/z_R) \sim w_0$.
By taking all the factors into account,
we find $\bar{N}_1 \sim {\cal O}(10^{22})$,
corresponding to $\sim 10$~kJ per pulse focusing.

%

If the one-beam focusing setup cannot detect any signals,
it excludes all of mass range below $m_{cut}$ and all of couplings
stronger than $gM^{-1}$ in that mass range. This is because, in principle, the
$\sin^{-4}\vartheta_r$ dependence in (\ref{kinm_13})
enhances the cross section for the entire lower mass range below $m_{cut}$.
We note that there is no known physical cut-off on the minimum $\vartheta$
in the photon-photon interaction. This by itself can be a subject
to be studied, since it is possibly related with the texture of
vacuum which may prevent the smooth propagation of photons in vacuum.
This may introduce a finite cut-off in the minimum incident angle
between two photons.
If the intensity discussed above is available, we may be able to search for
extremely light fields in the totally unprobed domain: $m<10^{-10}$~eV
and $gM^{-1}>10^{-23}$~GeV${}^{-1}$. This is a remarkable improvement compared
to the axion (pseudoscalar field) searches so far taken. They provided the
upper limit in the domain: $m>10^{-6}$~eV and $gM^{-1}>10^{-13}$~GeV${}^{-1}$
~\cite{PDG}.
The way to estimate the mass-coupling limit in our suggestion
should be contrasted to the conventional ones \cite{PDG}
where experiments provided upper bounds on the mass-coupling limit.
The gap between the two domains, however, can be filled by changing
the experimental parameters in our concept.
Even if there is no signals in the one-beam focusing, we only have to
update the condition so that it satisfies $\Delta\vartheta > \vartheta_r$
for heavier masses by increasing $\Delta\vartheta$.
In such heavier mass region, however,
two-beam crossing geometry relaxes the constraints on the optical design
such as focal length.
In either case the one-beam focusing setup considered in this paper
provides a basis to define the mass-coupling limit as well as
the necessary beam intensity as we have demonstrated here.

%
%
%
%
Our arguments above have been based on the approach
in which each interaction arise incoherently yielding the observed result by
$\bar{N}^2\overline{|{\cal M}|^2}$.  It is worth noting, however, the
interaction may occur coherently because the light field exchange gives
a sufficiently long-range interaction, and hence individual scattering
centers in the  beam are indistinguishable. This coherence
effect results in $\bar{N}|\bar{N}\overline{{\cal M}}|^2$ which is enhanced
by a factor of $\bar{N}$ over the yield expected from the incoherent summation,
as was discussed in the long-wavelength neutrino interaction
on a bulk target \cite{SMITH_COHERENT}.
Upon more careful analysis of our system the question may be addressed
if the required laser intensity can be relaxed due to a collective frequency
shift rather than the shift on the single-photon basis.

%
%
A major instrumental background for the frequency doubled photon
is in principle expected to be the second harmonic generation~(SHG)
from the final focusing optical element. The dominant source of SHG
may be the interface between the residual gas and the surface of the
optical element where the centrosymmetry is maximally broken.
Even from the maximal estimate $\sim10^{13}$W/cm${}^2$ for a typical damage
threshold, we expect a negligible amount of
$10^{-10}$ SHG photons from a 1m${}^2$ aperture
size with a 10~fs irradiation, if the optical components are housed
in the vacuum containing $10^{10}$~atoms/cm${}^3$
($\sim 10^{-5}$~Pa)~\cite{SHG}.
The confirmation of the negligible amount of the background SHG
is a crucial subject for the present concept.

%
%
As a dominant physical background we expect the lowest-order QED
photon-photon scattering with the forward cross section  $\sim
(\alpha^2/m_e^4)^2 \omega^6 \vartheta^4$ \cite{XsecQED}.
This turns out to be much smaller than \reflef{eq_dSdO3}) due to
the distinct behavior with respect to the incident angle $\vartheta$.
This indicates the lowest-order QED contribution is negligible.

\begin{figure}
\includegraphics[width=1.0\linewidth]{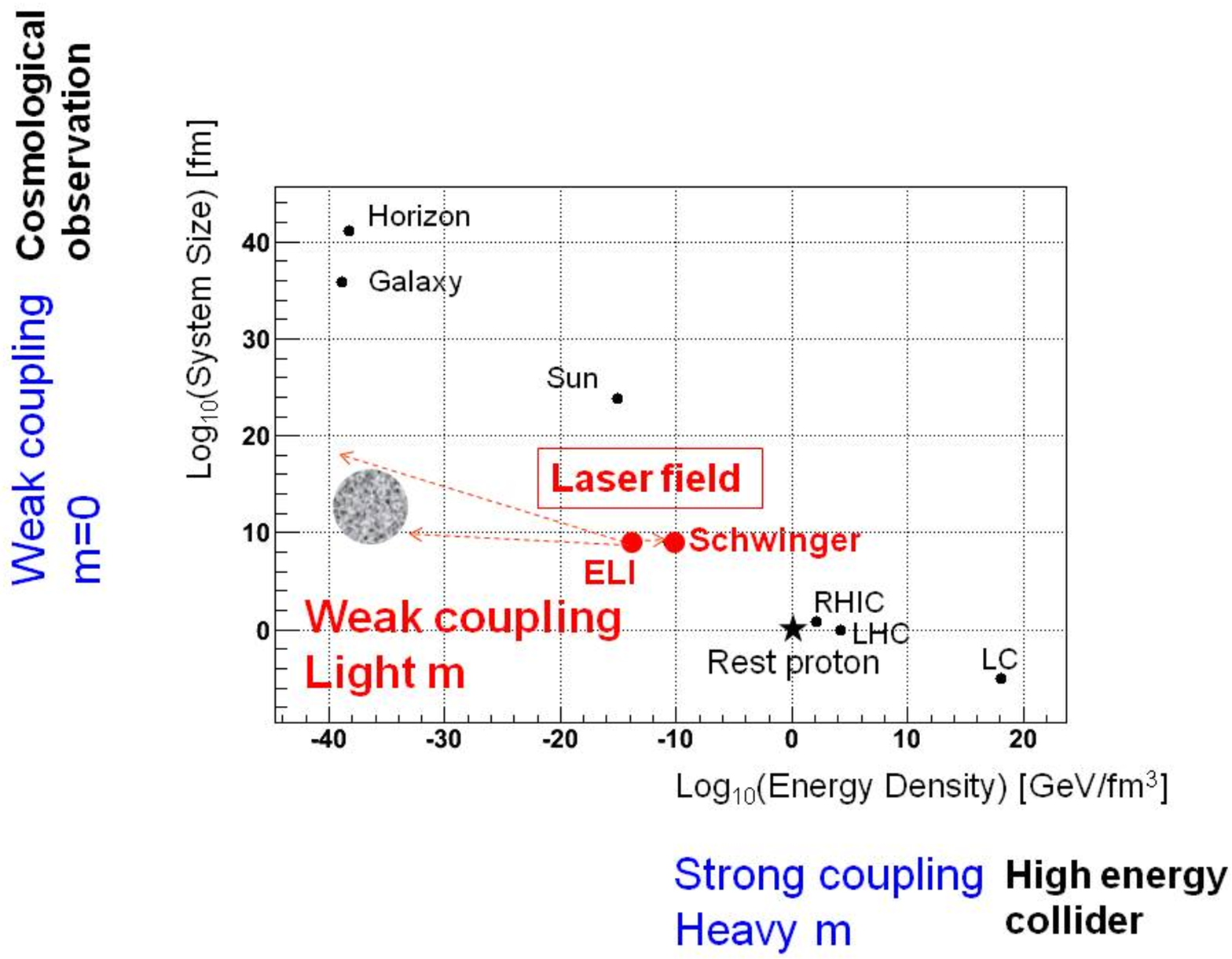}
\caption{
Experimental ballparks of various approaches to probe matter and vacuum
in the system size vs. the energy density.
Selected systems are
LC: electron-positron collision in the center of mass energy $E_{cms}=1$TeV at
the future linear collider~\cite{Colliders} assuming
the electron size $10^{-18}$cm,
LHC: proton-proton collision in $E_{cms}=14$TeV
at Large Hadron Collider~\cite{Colliders},
RHIC: gold-gold collision in $E_{cms}=200$GeV per nucleon pair at
Relativistic Heavy Ion Collider~\cite{Colliders}, the rest proton
indicated by the asterisk as the origin of this plot,
ELI: an optical laser pulse expected in ELI project~\cite{ELI},
Schwinger: Schwinger limit~\cite{Schwinger},
Sun, the Milky Way Galaxy and event horizon with $\Omega_{tot} \sim 1.0$ and
$h \equiv H_0$/100[km/s/Mpc]$\sim 0.7$~\cite{Cosmology}.
The energy density axis is qualitatively interpreted as the inverse of
the force range or the mass scale $m$ of exchanged force, because
the mean free path becomes shorter in higher density states as
long as the coupling to matter is not weak.
On the other hand the coupling strength to matter $gM^{-1}$ defined in
(\ref{eq_phisigma}) qualitatively reflects the necessary size of matter
or vacuum in order to make the interaction manifest.
The arrow to the higher energy
density toward Schwinger limit is the direction to probe nonlinear QED
interactions and also understanding of the non-perturbative nature of
the intense field, while the arrows directing lower energy density region
indicate the extensible domain by using co-propagating intense laser fields,
since the mass range may be covered from $\sim 1$~eV to $\sim 0.1$~neV
and the coupling may be probed as weak as gravitational coupling
for lighter mass scales. The energy density in this direction depends on
the context. In the context of the scalar field as a candidate of
dark energy in \cite{DEreview, DEsearch},
the energy density should be close to that of the event horizon.
}
\label{Fig9}
\end{figure}

\section{Conclusion}\label{sec7}
We have suggested an approach to probe the nature of vacuum with intense lasers.
Two main methods have been explored. The first is
the phase contrast Fourier imaging at the focal plane
to measure the phase shift of propagating light under the
intense laser field. The second is the extremely light resonance search
via higher harmonic generation by focusing a single intense laser.
Both are based on similar ideas extended from those already developed
to probe matter.

In these methods we take note of the nonlinearities of
vacuum that are either considered to exist but at an extremely minute level
or speculated to wait for our sensitive detection of an extraordinarily
feeble signal. In order to detect these weak nonlinearities, we need to
spectacularly enhance the signal. The large leap in enhancing these signals is
garnered by the combination of (1) the rapid development of intense laser
technology and its adoption here; (2) the employment of our suggested
techniques that circumvent potentially huge noise and allow the enhanced
interaction with the pursued fields. The former element (1) may be brought in,
for example, by the intense optical laser beyond kJ. For the latter factor (2)
we have suggested for the nonlinear QED problem a method of avoiding the
pedestal noise in our phase contrast Fourier imaging. 
For the exploration of possible
new low mass fields we have suggested a method to hit a resonance with
photons co-propagating over a long distance.

With the phase contrast Fourier imaging
the nonlinear QED effects of the Euler-Heisenberg Lagrangian
should be detected no more than the energy of laser around 10J.
The method provides a window for scoping vacuum via the dynamics of
electron mass scale virtually.
Such detection has never been made to date, and it heralds the research in
the physics of vacuum with the high field approach. 
With the detection of second harmonic generation in the co-propagating setup
we should be able to survey a large sweep of the energy domain
(eV to $10^{-10}$eV) of the intermediating vacuum fields.
If and when we pick up some signal in one particular energy range,
perhaps we can zoom in to this specific energy
(and thus wavelength) of photons by arranging the various knobs such as the
crossing angle and the (long) beat wavelength of the electromagnetic waves by
setting up a specific resonance cavity, we may be able to further increase the
sensitivity and more deeply study their properties.

Given the high intense optical laser beyond $1$kJ per fs-pulse duration
in the near future, the realization of these suggestions may become
an exciting challenge for future experiments of explorations of
the physics of vacuum.

Figure \ref{Fig9} illustrates the experimental ballparks of various approaches
to probe matter and vacuum, in terms of the system size as a function of
the energy density.
The energy density axis is qualitatively interpreted as the inverse of
the force range or the mass scale $m$ of exchanged force in (\ref{mxelm_7}),
because the mean free path becomes shorter in higher density state as
long as the coupling to matter is relatively strong.
On the other hand, the coupling to matter $gM^{-1}$ in 
(\ref{mxelm_7}) or (\ref{eq_phisigma})
qualitatively reflects the necessary size of matter or vacuum in order
to make the interaction visible.
The Galileo type telescope observes gravitational
phenomena. These are on the extremely
weak coupling of $M_P^{-1}$ with zero mass exchange.
High energy particle colliders, the Rutherford type microscopes focus on
particle generation phenomena. These are due to
strong couplings with heavy mass exchanges within fm scale.
There is a huge gap between these two approaches. In other words the region of
weak couplings with finite but light mass exchanges have been hardly probed
so far. It is quite natural to start exploring if there exist
important pieces of the puzzle of nature in these domains.
These explorations might allows us deeper
understanding of the nature of vacuum such as dark energy~\cite{DE}.
The progress of modern physics has been simply driven by
those two types of the experimental approaches.
The proposed method with high intense laser
probes semi-macroscopic vacuum compared to particle physics and
much smaller scale of vacuum compared to cosmology.
Provided such semi-macroscopic vacuum scope,
we increase our observational window into a new parameter regime of vacuum.

\begin{acknowledgments}
The research has been supported by the DFG Cluster of Excellence MAP
(Munich-Center for Advanced Photonics).
K.Homma acknowledges
for support by the Grant-in-Aid for Scientific Research no.21654035
from MEXT of Japan in part.
T. Tajima is the Blaise Pascal Chair of Ecole Normale Superieure.
We thank advice, collaboration, and encouragements from our colleagues,
including : G. Mourou, Y. Fujii, H. Gies, R. Sch\"{u}tzhold, G. Dunne,
M. Marklund, P. Shulkla, S. Bulanov, T. Esirkepov, C. Keitel, M. Siemko,
R. Assmann, F. Krausz, S. Karsch, M. Zepf, S. Sakabe,
K. Kondo, K. Fujii, T. Takahashi,
A. Suzuki, F. Takasaki, J. Urakawa, S. Ushioda, J. Rafelski, W. Sandner,
T. Heinzl, N. Naumova, and P. Chen.
\end{acknowledgments}



\begin{references}
\bibitem{footnoteBW}
The complicated analytic solution of the integral
is found. We have checked the behavior of the solution around the width $a$.
\end{references}
\end{document}